\documentclass[preprint,aps,prc,nofootinbib,showpacs]{revtex4}

\usepackage{graphicx}
\usepackage{graphics}
\usepackage{epsfig}

\begin{document}
\title{Pseudoscalar Mesons in Hot, Dense Matter}
\author{P. Costa}
\email{pcosta@teor.fis.uc.pt}
\author{M. C. Ruivo}
\email{maria@teor.fis.uc.pt}
\affiliation{Departamento de F\'{\i}sica, Universidade de Coimbra,
P-3004-516 Coimbra, Portugal} 
\author{Yu. L. Kalinovsky\footnote{Permanent Address: Laboratory of Information Technologies,
Joint Institute for Nuclear Research, Dubna, Russia}}
\email{kalinov@qcd.phys.ulg.ac.be}
\affiliation{Universit\'{e} de Li\`{e}ge, D\'{e}partment de Physique B5, Sart Tilman, 
B-4000, LIEGE 1, Belgium}
\author{C. A. de Sousa}
\email{celia@teor.fis.uc.pt}
\affiliation{Departamento de F\'{\i}sica, Universidade de Coimbra,
P-3004-516 Coimbra, Portugal} 
\date{\today}
\begin{abstract}
Phase transitions in hot and dense matter  and the in--medium behavior of  pseudoscalar mesons 
($\pi^{\pm}\,, \pi^0\,,K^{\pm}\,, K^0\,,\bar K^0,\eta\,\,\mbox{and}\,\,\eta' $) are investigated, in the framework of the three flavor Nambu--Jona-Lasinio model, including the 't Hooft interaction, which breaks the $U_A(1)$ symmetry.
Three different scenarios are considered: zero density and finite temperature, zero temperature and finite density in quark matter with different degrees of strangeness, and finite temperature and density.
At $T=0$, the role of strange valence quarks in the medium is discussed, in connection with the phase transition and the mesonic behavior. It is found that the appearance of strange quarks, above certain densities, leads to meaningful changes in different observables, especially in matter with $\beta$ --equilibrium.
The behavior of mesons in the $T-\rho$ plane is analyzed in connection with possible signatures of restoration of symmetries.  
\end{abstract}
\maketitle


\section{Introduction}

Understanding the  behavior of matter under extreme conditions is nowadays a challenge in the physics of strong interactions.   
Different regions of the QCD phase diagram are object of  interest and major theoretical and experimental efforts have been dedicated to the  physics of relativistic heavy--ion collisions, looking for signatures of the quark gluon plasma [QGP] \cite{kanaya,rhic,cern}. 
Special attention has also been given to neutron stars which are a natural laboratory to study matter at high-densities.
 
There are indications from Lattice QCD that the transition from the hadronic phase to the quark gluon plasma   is probably associated to the transition from the Nambu--Goldstone realization of chiral symmetry to the Wigner--Weyl phase.
While  the phase transition with finite chemical potential and zero temperature is expected to be first-order, at zero chemical potential and finite temperature there will be a smooth crossover.  
Experimental and theoretical efforts have been done  in order to explore the $\mu-T$ phase boundary. 
Recent Lattice results indicate a critical "endpoint", connecting the  first-order phase transition with the crossover region, at $T_E=160\pm35\mbox{MeV},\,\mu_E=725\pm35\mbox{MeV}$ \cite{Fodor}. 
Understanding the results of experiments at BNL \cite{rhic} and  CERN \cite{cern} provides a natural  motivation for these studies.   

The rich content of the QCD phase diagram has been recently explored in the direction of   high-density and cold matter, that can exist in neutron stars, where a "color-flavor locking" [CFL] phase, exhibiting a variety of interesting physics is expected to occur \cite{CFL,rajagopal}.       

Restoration of symmetries and deconfinement are expected to occur at high-density and /or temperature. In this concern, the study of observables of pseudoscalar mesons is particularly important. Since the origin of these mesons is associated to  phenomena of spontaneous and explicit symmetry breaking, its  in--medium behavior  is expected  to  carry relevant signs for possible restoration of symmetries. On the classical level and in the chiral limit, the QCD Lagrangian has two chiral symmetries, the $SU(N_f)$ and the $U_A(1)$. This  would imply the existence of nine Goldstone bosons, for $N_f=3$, and,  in order to have  mesons with finite mass, chiral symmetry must be explicitly broken {\it ab initio} by giving current masses to the quarks. However, in nature there are only eight light pseudoscalar mesons, the octet ($\pi,\, K,\, \eta$);  the $\eta'$ has a mass too large to be considered a remnant of a Goldstone boson, so its mass must have a different origin. In fact, the $U_A(1)$ symmetry is not a real symmetry of QCD, since it is broken at the quantum level, as pointed out by Weinberg \cite{Weinberg}. The breaking of the $U_A(1)$ symmetry can be described semiclassically by instantons, which has the effect of giving a mass to $\eta'$ of about 1 GeV. On the other side, this $U_A(1)$ anomaly causes flavor mixing, which has the effect of lifting the degeneracy between $\eta$ and $\pi^0$. So a percentage of the $\eta $ mass comes from the $U_A(1)$ symmetry breaking. 

So far as the restoration of symmetries is concerned, there are two possible scenarios \cite{shuryak}: only $SU(3)$ chiral symmetry is restored or both, $SU(3)$ and $U_A(1)$ symmetries, are restored. 
The behavior of $\eta'$ in medium or of related observables, like the topological susceptibility \cite{ohta1}, might help to decide between these scenarios.
A decrease of the $\eta'$ mass could lead to the increase of the  $\eta'$ production cross section, as compared to that for $pp$ collisions \cite{kapusta}.

Strange quark matter [SQM] has attracted a lot of interest since  the suggestion \cite{witten} that it could be the absolute ground state of matter. Stable SQM in $\beta$ --equilibrium is expected to exist in the interior of neutron stars or even be the constituent matter of  highly bound compact stars ("strange quark stars").  Lumps of SQM, the strangelets, might also be formed in earlier stage of heavy-ion collisions (in this case $\beta$ --equilibrium may not be achieved). Experiments of ultrarelativistic heavy-ion collisions at BNL and CERN are proposed to search for strangelets \cite{faessler7,castor}, but up to now there is no evidence of such objects \cite{faessler8}. Several  studies on the behavior of matter with different  strangeness fractions are known (see \cite{buballa,faessler} and references therein).

An interesting problem in flavor asymmetric matter is the behavior of the charge multiplets of mesons \cite{Kaplan,kubodera}. These charge multiplets, that are degenerated in vacuum or in symmetric matter ($\rho_u\,=\,\rho_d\,=\rho_s$), are expected to have a splitting in flavor asymmetric matter. In particular, the masses of kaons (antikaons) would increase (decrease) with density. A similar effect would occur for $\pi^-$ and $\pi^+$ in neutron matter. 
A slight raising of the $K^+$ mass and a  lowering of the $K^-$ mass \cite{RSP,waas,Cassing,ko,Schaffner} seems to be compatible  with the analysis of data on kaonic atoms \cite{Friedmann94} and with the results of KaoS and FOPI collaborations at GSI \cite{Herrman}.
The driving mechanism for the mass splitting is attributed mainly to the selective effects of the Pauli principle, although, in the case of $K^-$, the interaction with the $\Lambda(1405)$ resonance plays a significant role as well, in the low-density regime \cite{waas}.

Effective quark models are useful tools to explore   the behavior of 
matter at high-densities or temperatures. Nambu--Jona-Lasinio [NJL] \cite{njl}  type models have been extensively used
over the past years to describe low-energy features of hadrons and also to
investigate restoration of chiral symmetry with temperature or density
\cite{RSP,njl,RuivoSousa,SousaRuivo,Hiller,maria,buballa,kuni,njlT,ebert,klev2,bla,costa}.

The NJL model is an effective quark model where the gluonic degrees of freedom are supposed to be integrated out and, besides its simplicity, has the advantage of incorporating important symmetries of QCD, namely chiral symmetry.
Since the model has no confining mechanism, several drawbacks are well known.
It should be noticed, for instance, that the $\eta'$ mass is very close to the $\bar q q$ threshold and, depending on the parameterization, it can be above this threshold. In this case, this meson is described, even at zero temperature and density, as $\bar q q$ resonance, which would have the unphysical decay in $\bar q q$ pairs, and the definition of its mass is unsatisfactory. 

The behavior of $SU(3)$ pseudoscalar mesons in hot matter has been studied within the framework of NJL model in \cite{kuni,njlT,klev2,bla}. Different studies have been devoted to the behavior of pions and kaons at finite density in flavor symmetric \cite{RuivoSousa,SousaRuivo} or asymmetric matter \cite{SousaRuivo,maria,costa}.

A model aiming at describing hadronic behavior in the medium should account for the great variety of particle--hole excitations that the medium can exhibit, some of them with the same quantum numbers as the hadrons under study \cite{Cassing,ko}. 
Particle--hole excitations with the same quantum numbers of kaons have been discussed in \cite{kubodera}. It has been shown, within the framework of NJL models, that low-energy pseudoscalar modes, which are excitations of the Fermi sea, occur in flavor asymmetric media \cite{RuivoSousa,SousaRuivo,Hiller}.  
Such studies were carried out in quark matter simulating nuclear matter $(\rho_u = \rho_d\,, \rho_s=0)$,
for charged kaons,  and neutron matter without $\beta$ --equilibrium, for charged pions.
The role of the 't Hooft interaction was not taken into account in the last case.
The combined effect of density and temperature, as well as the effect of vector interaction, was discussed for the case of charged kaons \cite{RSP,costa}.
Only densities below $\sim 3 \rho_0$ were considered. The high-density
region ($\rho_n > \rho_n^{cr}=2.25 \rho_0$) of  quark matter  simulating neutron matter in weak equilibrium,  having in mind  the study of  the behavior of kaonic (charged and neutral) and pionic (charged) excitations was investigated in \cite{costaruivo}, and the behavior of neutral pseudoscalar mesons in hot and dense matter was investigated in \cite{neutral}.

This paper is devoted to investigating the phase transition in hot and dense matter, and the in--medium behavior of the pseudoscalar mesons, in the framework of the $SU(3)$ Nambu--Jona-Lasinio model, including the 't Hooft interaction, having in mind to look for manifestations of restoration of symmetries and to discuss the role of the strangeness degree of freedom.

We present the model and formalism in the vacuum (Section \ref{model}) and at finite density and temperature (Section \ref{att}). We will present and discuss our results for the phase transition  and meson behavior in three scenarios: finite temperature and zero density (Section \ref{phase}); zero temperature and finite density in quark matter with different degrees of strangeness, with and without $\beta$ --equilibrium (Section \ref{asymm}). Finally, the meson properties in hot and dense matter  are investigated in Section \ref{hotand}.
In Section \ref{conclusion} we draw our conclusions.


\section{Model and formalism}\label{model}

The NJL model with three quarks can be described by the Lagrangian
\begin{eqnarray} \label{lagr}
{\mathcal L} &=& \bar{q} \left( i \partial \cdot \gamma - \hat{m} \right) q
+ \frac{g_S}{2} \sum_{a=0}^{8}
\Bigl[ \left( \bar{q} \lambda^a q \right)^2+
\left( \bar{q} (i \gamma_5)\lambda^a q \right)^2
 \Bigr] \nonumber \\
&+& g_D \Bigl[ \mbox{det}\bigl[ \bar{q} (1+\gamma_5) q \bigr]
  +  \mbox{det}\bigl[ \bar{q} (1-\gamma_5) q \bigr]\Bigr] \, .
\end{eqnarray}
Here $q = (u,d,s)$ is the quark field with three flavors, $N_f=3$, and
three colors, $N_c=3$. $\lambda^a$ are the Gell--Mann matrices, a = $0,1,\ldots , 8$, ${ \lambda^0=\sqrt{\frac{2}{3}} \, {\bf I}}$.
The explicit symmetry breaking part in (\ref{lagr}) contains the current quark masses $\hat{m}=\mbox{diag}(m_u,m_d,m_s)$.
The last term in (\ref{lagr}) is the lowest six--quark dimensional operator and it has the $SU_L(3)\otimes SU_R(3)$ invariance but breaks the	$U_A(1)$ symmetry. This term is a reflection of the axial anomaly in QCD.
For general reviews on the three flavor version of the NJL model see \cite{njl,kuni,klev2}. 

Following a standard hadronization procedure, the following effective action is obtained:
\begin{eqnarray}\label{act}
  W_{eff}[\varphi,\sigma] &=&
  - \frac{1}{2} \left( \sigma^a S^{-1}_{ab}\sigma^b \right)
  - \frac{1}{2} \left( \varphi^a P^{-1}_{ab}\varphi^b \right)
  \nonumber \\ &&
  -i \mbox{Tr} \,  \mbox{ln} \Bigl[ i (\gamma_\mu \partial_\mu ) -
  \hat{m} + \sigma_a \lambda^a
  + (i \gamma_5 )(\varphi_a \lambda^a) \Bigr] \, .
\end{eqnarray}
The notation $\mbox{Tr}$ stands for the trace operation  over discrete indices ($N_f$ and $N_c$)  and integration over momentum.
The fields $\sigma^a$ and $\varphi^a$ are the scalar and pseudoscalar meson nonets and $S_{ab}\,, P_{ab}$ are  projectors defined in the Appendix A.

The first variation of the action (\ref{act}) leads to the gap equations,
\begin{eqnarray}\label{gap}
  M_i = m_i - 2g_{_S} <\bar{q_i}q_i> -2g_{_D}<\bar{q_j}q_j><\bar{q_k}q_k>\, ,
\end{eqnarray}
with $i,j,k =u,d,s$ cyclic. $M_i$ are the constituent quark masses and the quark condensates are given by: $<\bar{q}_i  q_i> =   -i \mbox{Tr}[ S_i(p)]\,\,, S_i(p)$ being the quark Green function.

To calculate the meson mass spectrum,  we expand the effective action
(\ref{act}) over meson fields. Keeping the pseudoscalar mesons only,
we have the effective meson action 
\begin{eqnarray}\label{act2}
W_{eff}^{(2)}[\varphi] =
-\frac{1}{2}\varphi^a \left[ P_{ab}^{-1} - \Pi_{ab} (P) \right] \varphi^b
= -\frac{1}{2}\varphi^a  D_{ab}^{-1}(P)  \varphi^b\,,
\end{eqnarray}
with $\Pi_{ab}(P)$ being the polarization operator (see Appendix A).
The expression in square brackets in (\ref{act2}) is the inverse nonnormalized meson propagator $ D_{ab}^{-1}(P)$.
The pseudoscalar meson masses are obtained  from the  condition
$(1- P_{ij} \Pi^{ij}(P_0=M,{\bf P}=0) )=0$. For the nondiagonal mesons
$\pi\,, K$, the polarization operator takes the form:
\begin{equation}
	\Pi^{ij} (P_0) = 4 \left((I_1^i + I_1^j)	-   [P_0^2-(M_i-M_j)^2]\,\,I_2^{ij}(P_0)\right),
\end{equation}
where  the  integrals are given in the Appendix A.

The quark--meson coupling constants are evaluated as \begin{eqnarray}\label{mesq}
	g_{M\overline{q}q}^{-2} = -\frac{1}{2 M} \frac{\partial}{\partial P_0} 			\left[\Pi^{ij}(P_0) \right]_{ \vert_{ P_0=M}} \, ,
\end{eqnarray}
where $M$ is the mass of the bound state containing quark flavors $i,j$.

To consider  the diagonal  mesons $\pi^0$, $\eta$ and $\eta'$ we  take into account the matrix structure of the propagator in (\ref{act2}). In the basis of $\pi^0 - \eta - \eta'$ system we write the projector $P_{ab}$ and the 
polarization operator   $\Pi_{ab}$ as  matrices:
\begin{eqnarray}\label{Pab}
	&&  {P}_{ab} =
  	\left(
		\begin{array}{ccc}
			P_{33} & P_{30} & P_{38} \\
			P_{03} & P_{00}& P_{08} \\
			P_{83} & P_{80}& P_{88}
		\end{array}
	\right)\,\,\,\,\,\,\mbox{and}\,\,\,\,\,\,{\Pi}_{ab} =
  \left(
	\begin{array}{ccc}
		\Pi_{33} & \Pi_{30} & \Pi_{38} \\
		\Pi_{03} & \Pi_{00}& \Pi_{08} \\
		\Pi_{83} & \Pi_{80}& \Pi_{88}
	\end{array}
  \right).
\end{eqnarray}

The nondiagonal matrix elements  
$P_{30}=\frac{1}{\sqrt{6}} g_D (<\bar{q}_{u}\,q_{u}> -<\bar{q}_{d}\,q_{d}>)$,
$ P_{38}=- \frac{1}{\sqrt{3}}g_D (<\bar{q}_{u}\,q_{u}>-<\bar{q}_{d}\,q_{d}>)$, 
$\Pi_{30}=\sqrt{2/3}[J_{uu}(P_0)-J_{dd}(P_0)]$ and $\Pi_{38}=1/\sqrt{3}[J_{uu}(P_0)-J_{dd}(P_0)]$
correspond to  $\pi^0 - \eta$ and $\pi^0 -\eta'$ mixing. In the case $<\bar{q}_{u}\,q_{u}>=<\bar{q}_{d}\,q_{d}>$, the $\pi^0$ is decoupled from the $\eta-\eta'$ and these elements vanish. The specific form of the nonvanishing elements of those matrices may be found in the Appendix A.

Defining the orthogonal matrix
\begin{eqnarray}
	{   O} =
	\left(
	\begin{array}{rr} 
		\mbox{cos}\theta  & -\mbox{sin}\theta \\
    \mbox{sin}\theta & \mbox{cos}\theta     
  \end{array}
	\right) \, ,
\end{eqnarray}
we may  find the $\eta - \eta'$ mixing angle $\theta$ via the condition to diagonalize $D_{ab}^{-1}(P)$ as ${   O}^{-1} D_{ab}^{-1}(P) {   O} =
\mbox{diag}(D^{-1}_{\eta}(P), D^{-1}_{\eta'}(P) )$.
To find the masses $M_\eta$ and $M_{\eta'}$, we use the inverse propagators
\begin{eqnarray}
	D_\eta^{-1} (P) =\left( { A}+{ C}\right) -  \sqrt{({ C}-{ A})^2+4{ B}^2} 
 	\\
	D_{\eta'}^{-1}(P) =\left( { A}+{ C}\right) +  \sqrt{({ C}-{ A})^2+4{ B}^2} 
\end{eqnarray}
with ${ A} = P_{88} -\Delta \Pi_{00}(P),
      { C} = P_{00} -\Delta \Pi_{88}(P),
      { B} = - (P_{08} +\Delta \Pi_{08}(P))$ and
     $\Delta = P_{00} P_{88}- P_{08}^2 $. 
In the rest frame, $ D_{\eta }^{-1}(P_0=M_{\eta },{\bf P }=0) =0,\,D_{\eta'}^{-1}(P_0=M_{\eta'},{\bf P }=0) =0$.
The mixing angle $\theta$ can be calculated by 
$\mbox{tan} 2 \theta = 2B/(A-C)$.
The coupling constants are determined by (\ref{mesq}). Note that from (\ref{mesq}) we have the quantities $g_{0\eta}, g_{8\eta}$ and $g_{0\eta'}, g_{8\eta'}$, from which we can obtain $g_{\eta(\eta')\bar{u}u}$ and $g_{\eta(\eta')\bar{s}s}$, using standard expressions (for more details see for example \cite{klev2,hadron03}).

The model is fixed by the coupling constants $g_S, g_D$ in the Lagrangian (\ref{lagr}), the cutoff parameter $\Lambda$ which regularizes momentum space integrals $I_1^{i}$ and $I_2^{ij}(P)$ and the current quark masses $m_i$.
For our numerical calculations we use the parameter set \cite{klev2}:\\
$m_u = m_d =        5.5$ MeV,
$m_s =            140.7$ MeV,
$g_S \Lambda^2 =   3.67$,
$g_D \Lambda^5 = -12.36$ and
$\Lambda =        602.3$ MeV,
\\
that has been determined by fixing the values\\
$M_{\pi} = 135.0$ MeV,
$M_K   = 497.7$ MeV,
$f_\pi =  92.4$ MeV and
$M_{\eta'}= 960.8$ MeV.
We also have \\
$M_{\eta}= 514.8$ MeV, $\theta (M_{\eta}^2) = -5.8^{\circ}$, $ g_{\eta \bar{u}u} =   2.40$,
$g_{\eta \bar{s}s} = -3.91$, \\
$\theta (M_{\eta'}^2) = -43.6^{\circ}$, $ g_{\eta'\bar{u}u} =  2.69$,
$g_{\eta' \bar{s}s} = -0.54$.
For the quark condensates we have:\\ 
$<\bar{q}_{u}\,q_u> = <\bar{q}_{d}\,q_d> = - (241.9 \mbox{ MeV})^3$ and
$<\bar{q}_{s}\,q_s> = - (257.7 \mbox{ MeV})^3$, and $M_u= M_d= 367.7\mbox{ MeV},\,M_s= 549.5\mbox{ MeV}$, 
for the constituent quark masses.


\section{The  model at finite temperature and chemical potential}\label{att}

Now we generalize  the NJL model to the finite temperature and chemical potential case. 
It can be done  by  the substitution (see \cite{kapustabook})
\begin{eqnarray}\label{subs}
  \int \frac{d^4 p}{(2\pi)^4} \longrightarrow
  \frac{1}{-i \beta} \int \frac{d^3 {\bf p}}{(2\pi)^3} \sum_n \, ,
\end{eqnarray}
where $\beta = 1/T$, $T$ is the temperature, and the sum is done over Matsubara frequencies $\omega_n=(2n+1)\pi T$, $n=0,\pm 1, \pm 2, \ldots$, so that $p_0 \longrightarrow i\omega_n + \mu$ with a chemical potential $\mu$.
Instead of integration over $p_0$, we have now the sum over Matsubara frequencies which can be evaluated as

\begin{eqnarray}
	-  \frac{1}{\beta} \sum_n h(\omega_n) &=&
	\sum_{\mbox{Re} z_m \neq 0} \biggl[\left(1-f(z_m) \right) \mbox{Res} [ 			h(\omega_n),z_m ]
	\nonumber \\
	&+& \bar{f}(z_m) \mbox{Res} [ \bar{h}(\omega_n),z_m ] \biggr] \, ,
\end{eqnarray}
where $f(z)$ and $\bar{f}(z)$ are the Fermi distribution functions for quarks and antiquarks,
\begin{eqnarray}
 	f(z) = \frac{1}{1+ e^{\beta (z-u)}},\,\,\,\bar{f}(z) = \frac{1}{1+ 								e^{\beta (z+u)}}.
\end{eqnarray}
As $1-\bar{f}(z)=f(-z)$, we introduce, for convenience, the Fermi distribution functions of the positive (negative) energy state of the $i$th quark:
\begin{eqnarray}\label{fermi}
	 n_i^{\pm}= f_i(\pm E_i) = \frac{1}{1+ e^{\pm\beta (E_i\mp\mu_i)}}.
\end{eqnarray}

The  integrals $I_1^i\,, I_2^{ij}(P)$ that enter in the expressions of the propagators depends now on the temperature $T$ and on the chemical potentials, in a standard way (see Appendix A). Having these integrals,
we can investigate the phase transition and meson properties in hot and dense matter.
First, we analyze the mesonic behavior at finite temperature and vanishing chemical potentials. 
Although  such a study was already performed in \cite{kuni,klev2}, it is pertinent to present the results here for the sake of comparison with our new findings at finite temperature and density, to be presented in next sections.

Fig. 1 shows the temperature dependence of $\pi$, $K$, $\eta$ and $\eta'$  meson masses, as well as  of $2 M_u$ and $M_u+M_s$ at $\mu = 0$.
One can see that, at low temperature, the masses of mesons (except the $\eta'$--meson that is always unbound) are lower than the masses of their constituents. In this case the integrals $I_2^{ij}$  are real.
The crossing of the $\pi$ and $\eta$ lines with $2 M_u$ and the $K$ line with $M_u+M_s$ indicates the respective Mott transition temperature  $T_M$  for these mesons. 
Mott transition comes from the fact that mesons are not elementary objects but are composed states of quarks, and is defined by the transition from a bound state to a resonance in the continuum of unbound states.
Above the Mott temperature we have taken into account the imaginary parts of the integrals $I_2^{ij}$ and used a finite width approximation \cite{klev2,bla}.
For our set of parameters, one can see that Mott temperatures for $\eta$ and $\pi$ mesons are: $T_{M\, \eta}=180$ MeV and $T_{M\, \pi}=212$ MeV. 
The $\pi$ and $K$ mesons become unbound at approximately the same temperature.

\begin{figure}[t]
	\begin{center}
		\epsfig{file=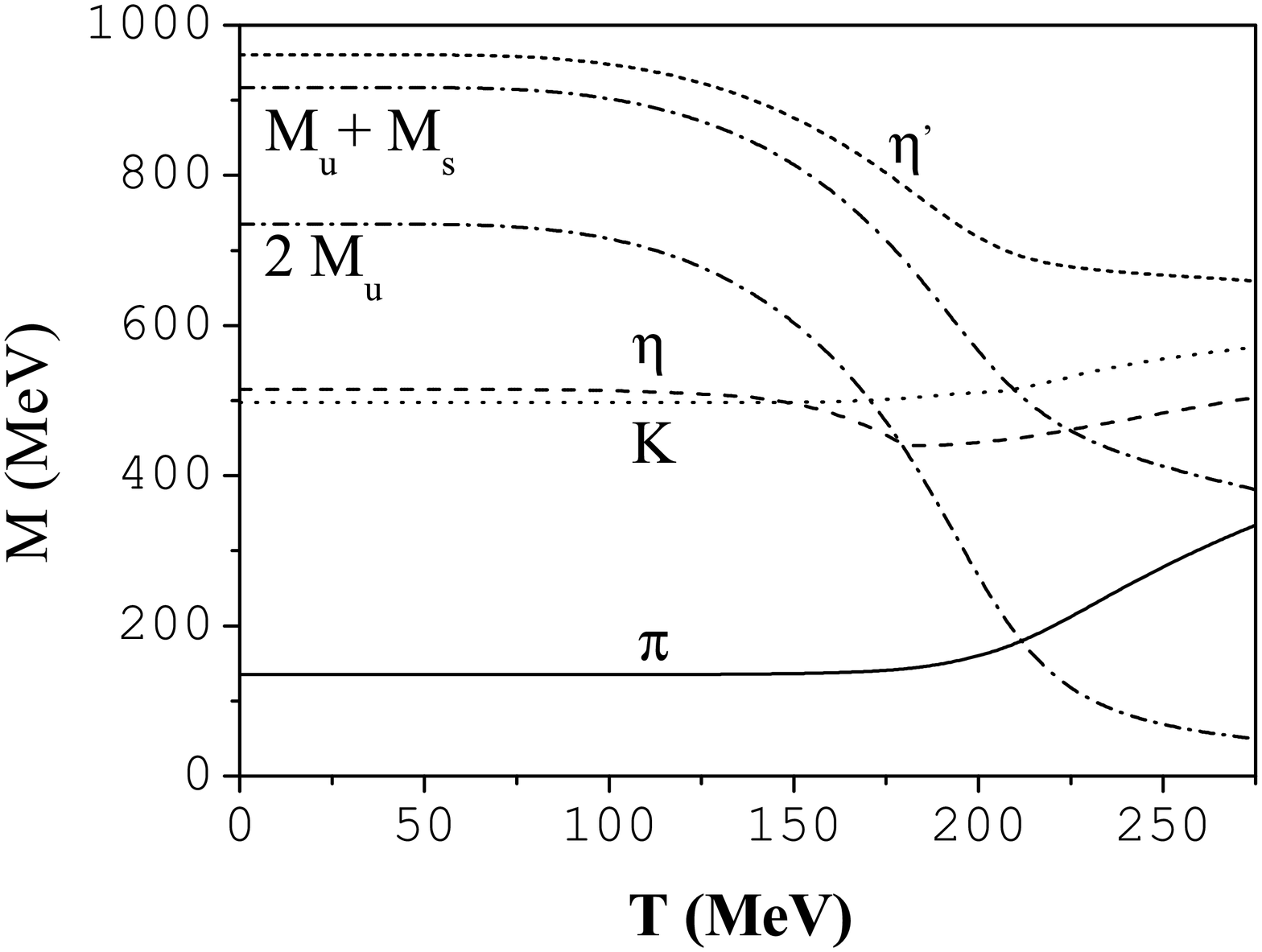, width=10.5cm,height=7.5cm}
	\end{center}
	\caption{Temperature dependence of the pion (solid line), kaon (dotted 							line), $\eta$ (dashed line) and $\eta'$ (short--dashed line) 								masses. The curves $2 M_u$ and $M_u +M_s$ (dot--dashed line) show 					the temperature dependence of the quark thresholds. The respective 					 Mott temperatures are: $T_{M \pi}\simeq T_{M K}=212$ MeV and 
					$T_{M \eta}=180$ MeV.}
\end{figure}


\section{The phase transition}\label{phase}

The nature of the phase transition in NJL type models at finite T and/ or $\mu$ has been discussed by different authors \cite{RSP,buballa,njlT,costaruivo,asakawa}. At zero density and finite temperature there is a smooth crossover, at nonzero densities different situations can occur. We will discuss this problem by analyzing the behavior of the pressure and of the energy per particle as functions of the baryonic density.

The equilibrium state may be determined as the point where the thermodynamical potential takes the minimum with the quark condensates
$<\bar{q}_i q_i >$ as variational parameters. The baryonic thermodynamic potential  has the following form:
\begin{equation}\label{tpot}
	\Omega (\rho ,T)= E- TS - \sum_{i=u,d,s} \mu _{i} N_{i}\,,
\end{equation}
where $E,\,S\, \mbox{and}\, N_i$ are, respectively, the internal energy, the entropy and the number of particles of the $i$th quark, that are given by the following expressions:
\begin{eqnarray}\label{energy}
	E &=&- \frac{ N_c}{\pi^2} V\sum_{i=u,d,s}\left\{
	\int {\tt p}^2 d {\tt p} \, \frac{%
	{\tt p}^2 + m_{i} M_{i}}{E_{i}}\, (n_{i}^{-}-n_{i}^{+})\,\,
	\theta (\Lambda^{2}-{\tt p}^2)\right\}  \nonumber \\
	&&- g_{S} \sum_{i=u,d,s}\, (\langle \bar{q}_{i}q_{i}\rangle )^{2}
	- 2 g_{D} \langle \bar{u}u\rangle \langle \bar{d}d\rangle \langle 				   \bar{s}s\rangle \,,
\end{eqnarray}
\begin{eqnarray}\label{entropy}
	S &=&-\frac{ N_c}{\pi^2} V \sum_{i=u,d,s}
	\int {\tt p}^2 d {\tt p} \,\, \theta (\Lambda^{2}-{\tt p}^2)
	\nonumber \\
	&&\times
	\biggl\{ \bigl[ n_{i}^{+} \ln n_{i}^{+}+(1-n_{i}^{+})\ln (1-n_{i}^{+})%
	\bigr] +\bigl[ n_{i}^{+}\rightarrow n_{i}^{-} \bigr] \biggr\} \, ,
\end{eqnarray}

\begin{equation}\label{np}
	N_i = \frac{ N_c}{\pi^2} V \int {\tt p}^2 d {\tt p}
	\left( n^-_i +n^+_i -1 \right) \theta (\Lambda^2-{\tt p}^2) \,.
\end{equation}
V is the volume of system and the quark density is determined by
the relation $\rho_i = N_i / V$. 

We define the pressure and the energy density such that their values are zero in the vacuum: 
\begin{equation} \label{p}
	p (\rho, T) = - \frac{1}{V}\left[ \Omega(\rho, T) - \Omega(0, T) \right] ,
\end{equation}
\begin{equation}\label{e}
	e(\rho, T) = \frac{1}{V}\left[E(\rho, T)-E(0, T)\right].
\end{equation}


\subsection{Chiral phase transition at zero temperature}

We start by analyzing the behavior of quark matter at zero temperature and, in order to discuss the role of the strangeness degree of freedom, we will consider matter with and without $\beta$ --equilibrium and, in the last case, we assume different fractions of strange quarks. 

In "neutron" matter  in chemical equilibrium, maintained by weak  interactions and with charge neutrality, the following constraints 
on the chemical potentials and densities of quarks and electrons should be imposed:
\begin{equation}\label{che}
	\mu_{d}=\mu_{s}=\mu_{u}+\mu_{e},
\end{equation}

\begin{equation}\label{ro}
	\frac{2}{3}\rho_{u}-\frac{1}{3}(\rho_{d}+\rho_{s})-\rho_{e}=0,
\end{equation}
with 
\begin{equation}
	\rho_{i}=\frac{1}{\pi^{2}}(\mu_{i}^{2}-M_{i}^{2})^{3/2}\theta(%
	\mu_{i}^{2}-M_{i}^{2})\,  \mbox{and} \,\rho_{e}=\mu_{e}^{3}/3\pi^{2}.
\end{equation}
We should note that if the chemical potential of the electron exceeds the rest mass of the muon ($\mu_e > M_{\mu}$) it becomes energetically favorable for an electron to decay into a $\mu^{-}$ via the weak process 
$e^{-}\rightarrow \mu^{-}+\overline{\nu}_{\mu}+\nu_e$ and we can have a Fermi sea of degenerate negative muons. Consequently, the condition for charge  neutrality would be $\frac{2}{3}\rho_{u}-\frac{1}{3}(\rho_{d}+\rho_{s})-\rho_{e}-\rho_{\mu}=0.$
However we have found $\mu_{e}^{max}=95.7$ MeV $<M_{\mu}$. So, we can neglect the muon contribution.

The thermodynamic potential, pressure and energy defined in (\ref{p}) and (\ref{e}) have to be modified in order to include the contribution of electrons. This leads to the following expressions for the pressure and energy density (see \cite{buballa}):
\begin{equation}
P = - e(\rho, 0)\,+ \sum_{i=u,d,s} \mu_i \rho_i\,+\frac{{\mu_e}^4}{12\pi^2},
\end{equation}
\begin{equation}
{\mathcal{E}} =e(\rho, 0) \,+\frac{{\mu_e}^4}{4 \pi^2}.
\end{equation}

Therefore the zeros of the pressure give the following expression for the energy density:
\begin{equation}
\mathcal{E} =\sum_{i=u,d,s} \mu_i \rho_i\,+\mu_e \rho_e.
\end{equation}
Defining the baryonic matter density as $\rho_n= \frac{1}{3} (\rho_u+\rho_d+\rho_s)$ and using the conditions (\ref{che}) and (\ref{ro}) it is straightforward to show that the energy per baryon at the zeros of the pressure takes the form
\begin{equation}\label{cri}
\frac{\mathcal{E} }{\rho_n}= \mu_u+2\mu_d.
\end{equation}
The pressure has a zero at $\rho_n=0$, the energy per baryon at that point being $M_{u}+2 M_{d}$ (since in vacuum the masses of the quarks $u,d$ are equal we will from now on denote this quantity as $3 M_v$). If there is another zero of the pressure, at $\rho_n\not=0$, that corresponds to a minimum of the energy, the criterion for stability of the system at that point is $\mu_u+2\mu_d < 3 M_v$. Let's now analyze  our results for the pressure and energy per baryon, that are plotted in Fig. 2 a)-b) as functions of $\rho_n/\rho_0$, where  $\rho_0 = 0.17$ fm$^{-3}$ is the normal nuclear matter density.
The pressure has three zeros, respectively at $\rho_n=0\,,  0.45 \rho_0\,,2.25 \rho_0$, that correspond to extrema of the energy per particle. For $\rho_n < 0.2 \rho_0$ the pressure and compressibility are positive, so the system could exist in  uniform gas phase, but it will not survive indefinitely, since the zero density state is energetically favored; for $ 0.2 \rho_0<\rho_n < 0.45 \rho_0 $ the system is unstable since the compressibility is negative, in fact $\rho_n=0.45 \rho_0$ corresponds to a maximum of the energy per particle; for $ 0.45 \rho_0<\rho_n < 2.25 \rho_0$, the pressure is negative, and the third zero of the pressure, 
$\rho=2.25 \rho_0$, corresponds to an absolute minimum of the energy.  In fact, at that point $\mu_u+ 2\mu_d= 1099.4 \mbox{ MeV}$, lower then $3 M_v= 1102.9 \mbox{ MeV}$. Above $\rho_n=2.25 \rho_0$, that we define as $\rho_{cr}$, we have again a uniform gas phase. The phase transition described in this model is a first-order one  since  there is a region where the pressure and/or compressibility are negative and,  given that at the critical density $\mu_u+2\mu_d < 3 M_v$ (we notice that the satisfaction of this condition is dependent on the parameter choice), the behavior described allows the interpretation that the uniform gas phase will break up into stable droplets of nonstrange quarks with partially restored chiral symmetry and density $\rho_n^{cr} = 2.25 \rho_0$, surrounded by a nontrivial vacuum (see also \cite{RSP,buballa,costa,rajagopal}).

\begin{figure}[t]
	\begin{center}
		\epsfig{file=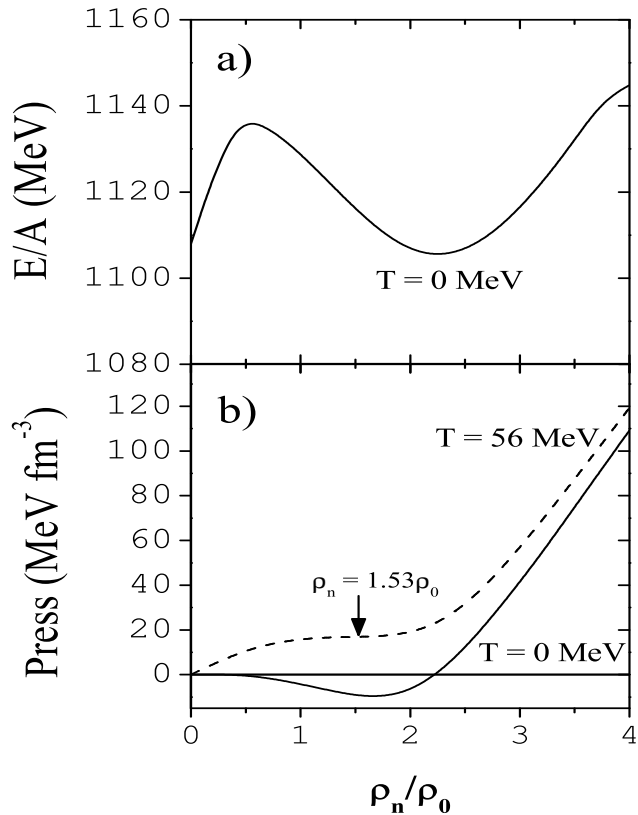, width=10.0cm,height=12cm}
	\end{center}
	\caption{Energy per baryon number  (a)) and pressure  (b)) as a function 		of density. Solid line: $T=0$, dashed line: $T=56$ MeV.}
\end{figure}

We will now discuss cases of matter without $\beta$-equilibrium. 
As we will show, the results are qualitatively similar to those discussed above, although some specific aspects deserve a closer analysis.
We consider three cases: Case I -- "neutron matter" without strangeness, ($\rho_d=2 \rho_u\,\,, \rho_s=0$); Case II -- matter with equal chemical potentials ($\mu=\mu_d=\mu_u=\mu_s$) with isospin symmetry, $\rho_u=\rho_d,\,\rho_s=\frac{1}{\pi^{2}}(\mu^{2}-M_{s}^{2})^{3/2}\theta(
\mu^{2}-M_{s}^{2})$; Case III --   matter entirely flavor symmetric ($\rho_d=\rho_u=\rho_s$). 

Let us make some comments concerning Case III, which intuitively seems the less natural scenario. Although rather schematic, this case simulates a situation where we can explore the hypothesis of absolutely stable SQM. It has been argued \cite{buballa,faessler} that SQM may only be stable if it has a large fraction of strange quarks ($\rho_s \approx \rho_u \approx\rho_d$).
The speculations on the stability of SQM are supported by the following observations: the weak decay of strange quarks into nonstrange quarks can be suppressed or even forbidden due to Pauli blocking, and, in addition, the inclusion of a new flavor degree of freedom allows for a larger decrease of strange quark mass which can produce a sizable binding energy. As we will see, Case III  confirms this tendency when compared with, for instance, Case I.

So, in spite of the simplicity of our assumptions, we think that the analysis of these types of matter   could  be useful as a guideline to understand   the role of the strangeness degree of freedom in matter  that is supposed to exist in neutron stars and in matter that might be formed in an early stage of heavy-ion collisions.  This will also be relevant for the discussion of the mesonic behavior that will be presented in next section.

We notice that in all cases there is an absolute minimum of the energy per particle at nonzero density and zero pressure, lower than  the vacuum constituent quark masses, so we have a first-order phase transition with the formation of quark droplets (see Fig. 3). However, this minimum of the energy is always higher than $930$ MeV, which is the energy per baryon number in atomic nuclei, and therefore we do not find absolutely stable quark matter. The lower energy per baryon is for Case II, and the higher binding energy, compared to the vacuum constituent quark masses, is for Case III. The minimum of the  energy occurs at $\rho_n=2.25 \rho_0$ in matter in $\beta$ --equilibrium and in Case I, at $\rho_n=2.33 \rho_0$ for Case II, and at $\rho_n=3.50\rho_0$ for Case III. Only in the last case we find stable SQM, since in the other cases the minimum occurs in the region where strange quarks do not exist (see Fig. 4, right panel). 

\begin{figure}[t]
\begin{center}
\epsfig{file=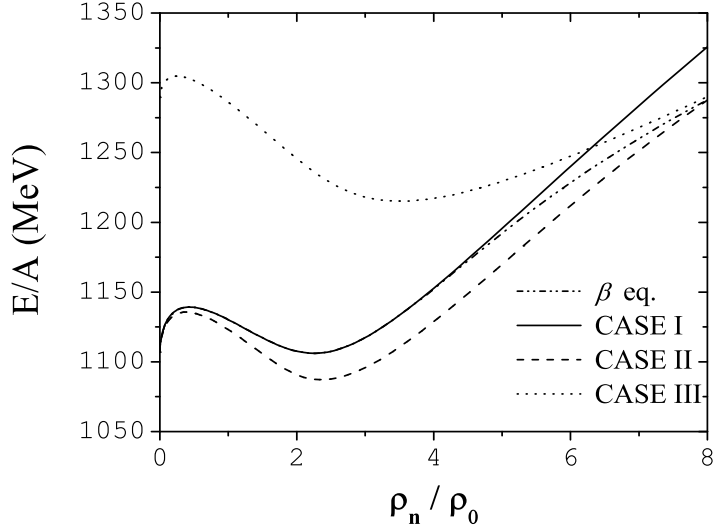, width=11cm,height=8.5cm}
\end{center}
\caption{Energy per baryon number for all types of quark matter considered at $T=0$.}
\end{figure}

Finally we add some considerations about the restoration of chiral symmetry. When chiral symmetry is broken {\em ab initio}, one can talk about restoration of chiral symmetry if the constituent quark masses drop to the current quark masses, indicating a transition from the chirally broken to approximately chirally symmetric phase. This would happen at asymptotic densities, when the Fermi momentum equals the cutoff. However, at densities considerably lower, one can see, or not, a clear tendency to the  restoration of chiral symmetry depending on the quark sector considered and on the strangeness fraction \cite{faessler}. To illustrate this point we plot in Fig. 4 the dynamical quark masses and chemical potentials as functions of $\rho_n/\rho_0$.
\begin{figure}[t]
\begin{center}
\hspace*{-0.2cm}\epsfig{file=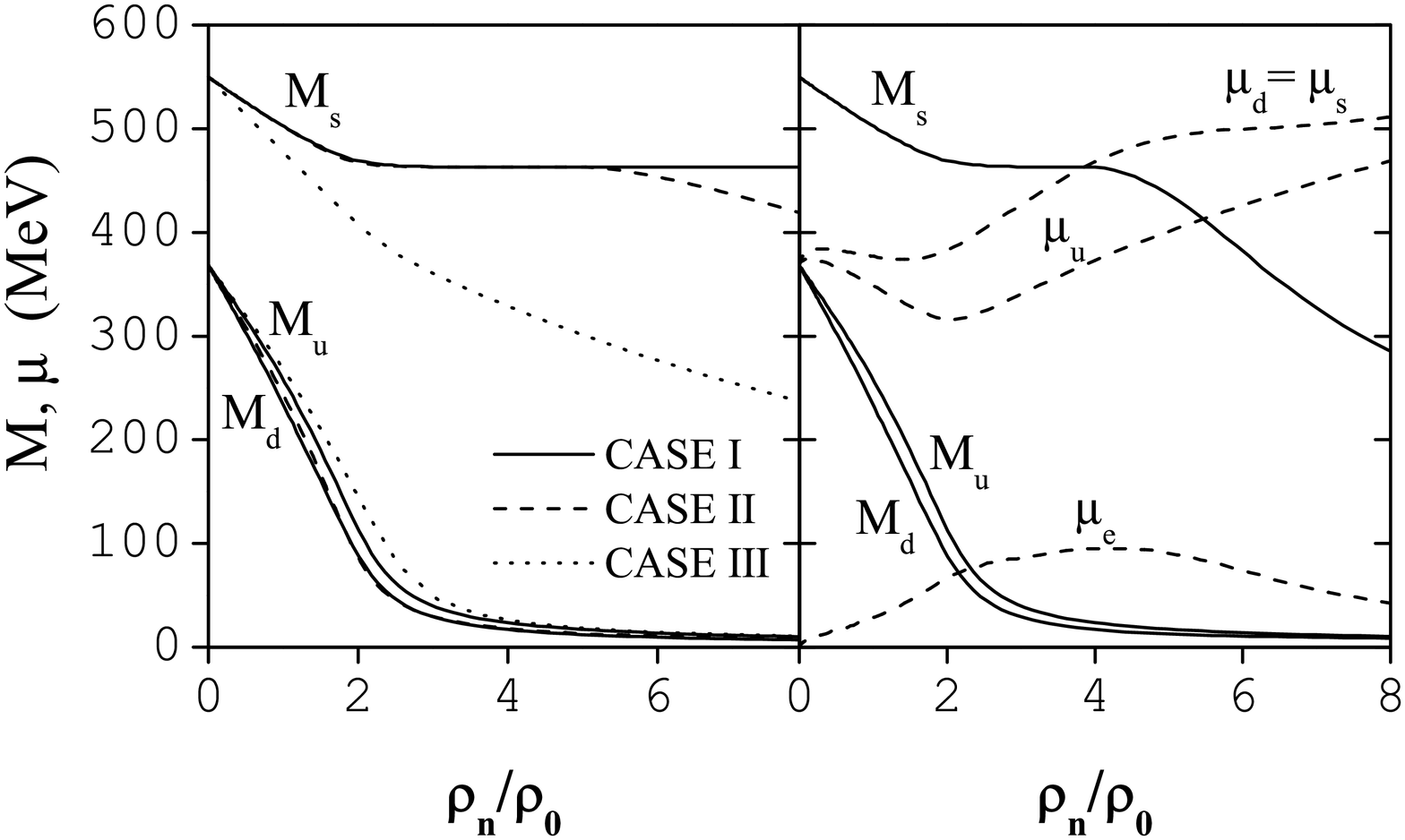, width=14.0cm,height=9cm}
\end{center}
\caption{Constituent quark masses $M_u$, $M_d$, $M_s$ and chemical potentials $\mu_u$, $\mu_d$, $\mu_e$ as functions of density at $T=0$.  Left panel: without $\beta$ --equilibrium ($M_u=M_d$ in Cases II and III). Right panel: with $\beta$ --equilibrium.}
\end{figure}
The nonstrange quark masses  decrease sharply in all cases, reflecting the quick restoration of chiral symmetry in this sector and, as expected, the behavior of the strange quark mass depends strongly on the amount of strange quarks in the medium. 
In all cases, except Case III, the mass decrease up to $\sim 2\rho_0$ is due to the 't Hooft contribution in the gap equations. 
In Case I it remains constant afterwords, since strangeness is put equal to zero by hand; in Case II and in $\beta$ --equilibrium, at densities above $5\rho_0$ and $3.8\rho_0$, respectively, the mass becomes lower than the chemical potential and a more pronounced decrease is observed. 
These densities are the onsets for the appearance of strange valence quarks, which become more important as the density increases.
In Case III, there are always strange valence quarks present, so the strange quark mass decreases more strongly, although, even in this case, is still away from the strange current quark mass.   
We observe that in matter in $\beta$ --equilibrium there is a tendency to the restoration of flavor symmetry, as the density increases: the chemical potentials $\mu_d=\mu_s$ and $\mu_u$ become closer and $\mu_e$ decreases.


\subsection{Chiral phase transition at finite temperature and density}

Now we discuss the phase transition in hot and dense matter and we consider only matter in $\beta $ --equilibrium (the other cases are qualitatively similar). Since for very low temperatures the absolute minimum of the energy turns to be at zero density, the phase transition is still first-order but the system is unstable against expansion. With increasing temperature, we will have a crossover above $T > T_{cl}= 56$ MeV. The point $T= 56\mbox{ MeV},\,\rho_n=1.53 \rho_0$, where the pressure is already positive and the compressibility has only one zero (see Fig. 2 b)), is identified, as usual, as  the critical endpoint, that connects the first-order phase transition and the crossover  regions. At this point we have a second order phase transition once the end point of a first-order line is a critical point of the second order. 
The values of the chemical potentials are: $\mu_u= 304.5$ MeV, $\mu_d=\mu_s= 353.3$ MeV and $\mu_n=(\mu_u+\mu_d+\mu_s)/3=337.0$ MeV.

In order to get a better insight on the nature of the phase transition in hot and dense matter, we plot in Fig. 5 a) the pressure in the $T - \rho$ plane. The region of the surface with negative curvature corresponds to the region of temperatures and densities where the phase transition is first-order.  
\begin{figure}
\begin{center}
  \begin{tabular}{cc}
    \hspace*{-0.5cm}\epsfig{file=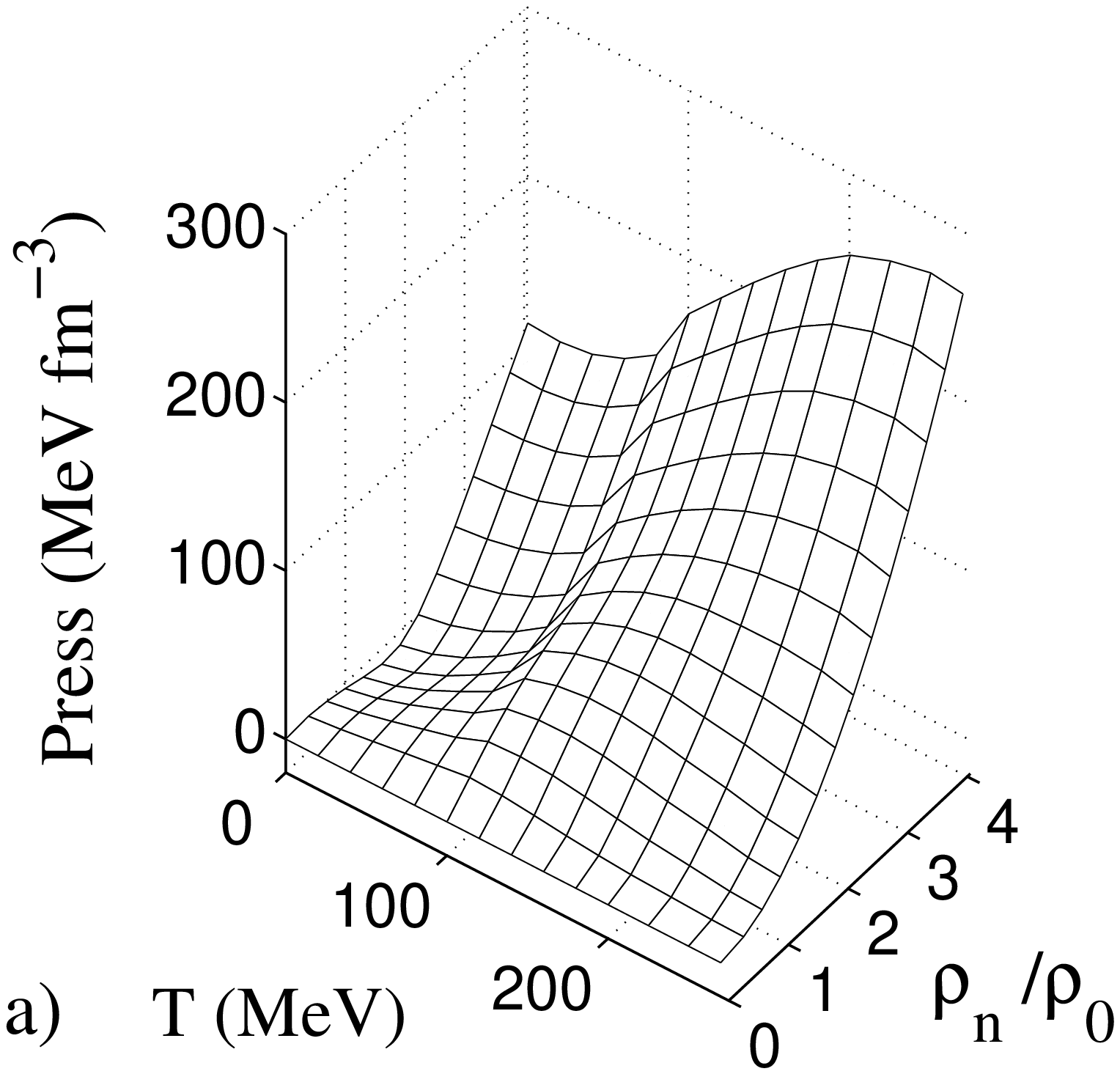,width=7.5cm,height=6.5cm} &
    \epsfig{file=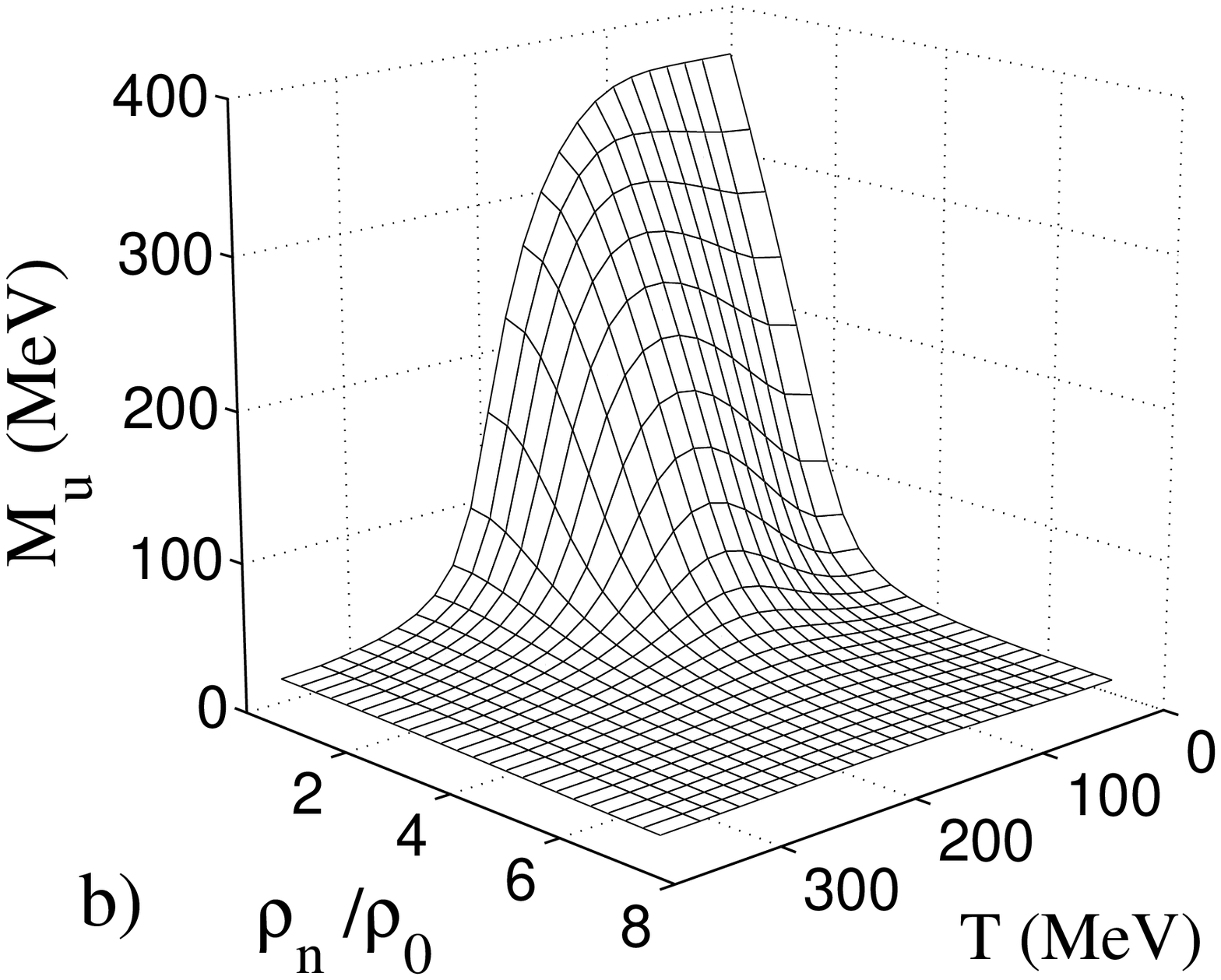,width=7.0cm,height=6.5cm} \\
   \end{tabular}
   \begin{center}
    \epsfig{file=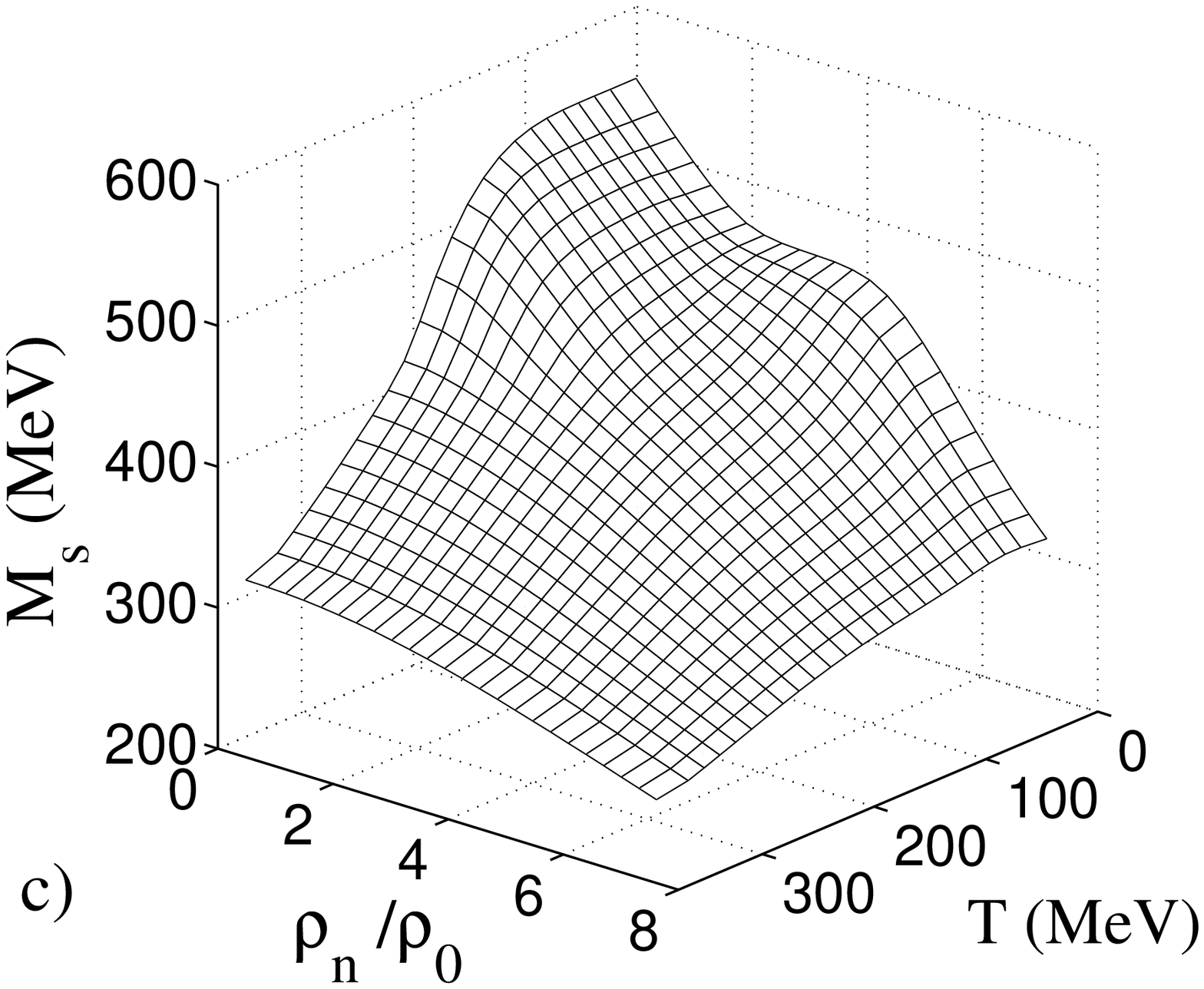,width=8cm,height=6.5cm}
   \end{center}
\end{center}
\caption{Combined effects of temperature ($T$) and density ($\rho_n/\rho_0$)
in neutron matter in $\beta$ --equilibrium: a) pressure; b) $M_u$; c) $M_s$.}
\end{figure}

We can also illustrate the degree of restoration of chiral symmetry in different flavor sectors by plotting  the constituent quark masses in the 
$T - \rho$ plane. In Fig. 5 b) (see also \cite{njlT}) one can see a clear manifestation of the restoration of chiral symmetry for the light quarks that is indicated by the pronounced flattening of the surface with increasing temperature and density. Fig. 5 c) shows a more smooth behavior of the strange quark, which reflects the well known fact that, as already discussed, chiral symmetry shows a slow tendency to get restored in the strange sector.


\section{Mesons in cold quark  matter}\label{asymm}

In this section we present our results  on the behavior of pseudoscalar mesons in   quark matter at $T=0$, giving special attention to matter in $\beta$ --equilibrium. The presence of strange valence quarks is related to a change in different observables. In order to discuss the relevance of the strange quark we will show the results for the mesonic behavior also in matter without $\beta $ --equilibrium.

One of our aims is to discuss what can in--medium pseudoscalar meson properties tell us about possible restoration of symmetries. If to discuss $U_A(1)$ symmetry it is enough to study neutral mesons, this is not sufficient if one wants to study also chiral symmetry, specially in the strange sector. Strangeness enters explicitly in the structure of $\eta, \eta'$ and of kaons, and influences the behavior of pions through the 't Hooft interaction. So, in order to have a comprehensive view  of restoration of chiral symmetry we have to include the study of pions and kaons. 


\subsection{Behavior of pions and kaons}\label{secpion}

Hadronic systems created in realistic heavy--ion collisions are strongly interacting. Many--body excitations may carry the same quantum numbers of the hadrons under investigation and, therefore, influence hadron properties in medium. Other many--body effects, such as the Pauli principle, lead to  modifications of hadronic properties, as well. How these effects are related with partial restoration of chiral symmetry is a challenging problem in many--body physics. As it will be shown, both effects are present in the behavior of the flavor multiplets of pions and kaons.

Let us analyze the results  for the masses of $K^+,K^-, K^0, \bar{K}^0$ and
for $\pi^+,\pi^0,\pi^-$, that are plotted as functions of $\rho_n/\rho_0$ in Figs. 6,7 and 9.

\begin{figure}[t]
\begin{center}
\hspace*{-0.2cm}\epsfig{file=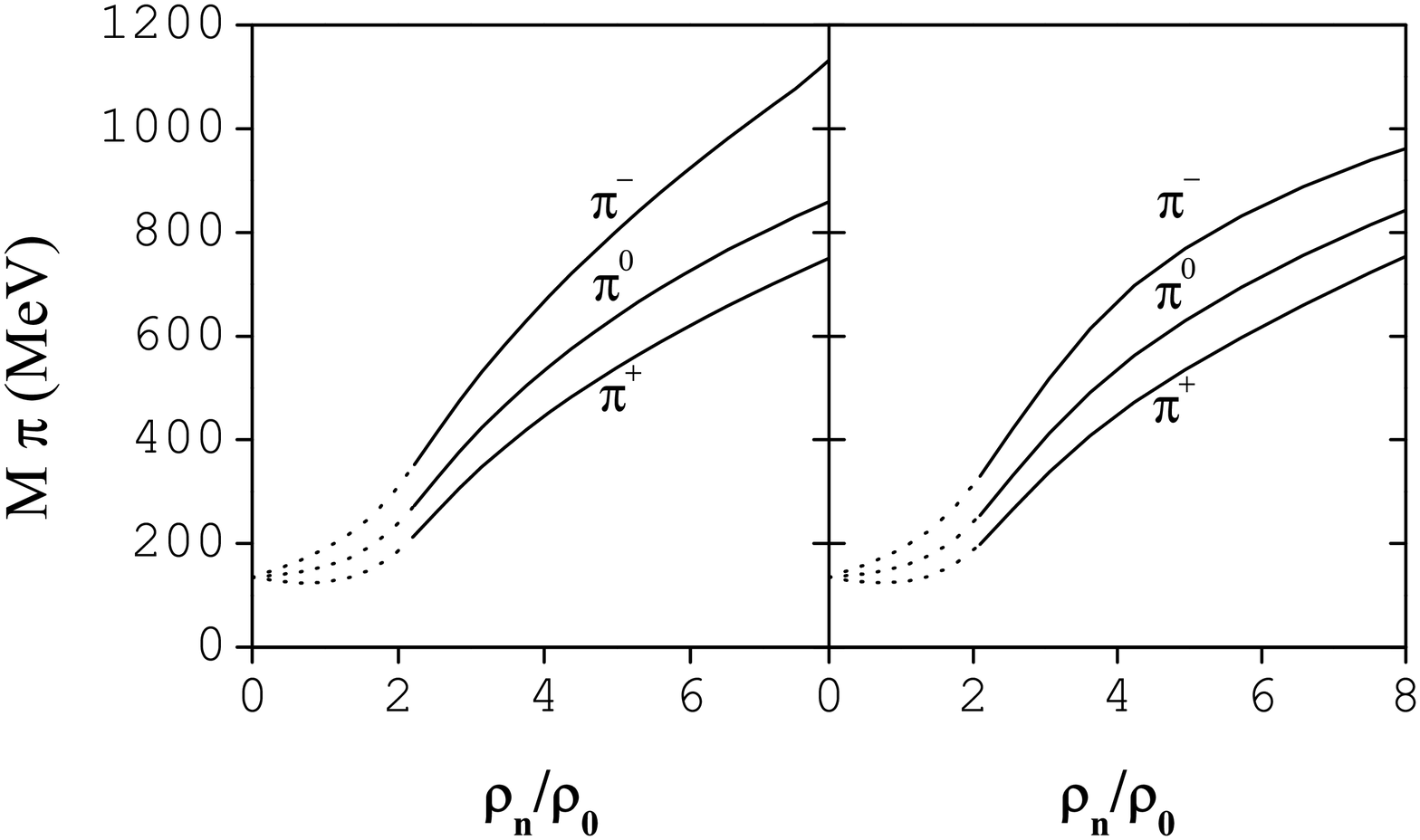, width=14.0cm,height=9cm}
\end{center}
\caption{
Pion masses as function of density without (left panel) and  
with (right panel) $\beta$ --equilibrium at $T=0$.}
\end{figure}

As it was already shown in other works \cite{RuivoSousa,SousaRuivo,costaruivo}, two kinds of solutions may be
found in asymmetric  matter for pionic  and kaonic modes  in NJL model,  corresponding respectively to excitations of the Dirac sea, that are modified by the presence of the medium, and excitations of the Fermi sea. Here, in order to appreciate the role of the strangeness degree of freedom, we start by comparing the results obtained in matter with $\beta$ --equilibrium and without, Case I. 

In order to get a better insight in the results, we plot  the limits of
the Dirac and of the Fermi sea continua of $\bar qq$ excitations with the quantum numbers of the mesons under study in Figs. 7 and 9 (dashed lines):
$\omega'= \sqrt{M^2_s + \lambda^2_s}+\sqrt{M^2_{u(d)}+ \lambda^2_s}$ 
is the lower limit of the Dirac continuum, and 
$\omega_{up}= \sqrt{M^2_s + \lambda^2_s}-\sqrt{{M^2}_{u(d)}+{\lambda^2}_s}$, 
$\omega_{low}= \sqrt{{M^2}_s+{\lambda^2}_{u(d)}}-\mu_{u(d)}$ are the upper and lower limits of the Fermi sea continuum with $\lambda_i$ the Fermi momentum of the quark $i$.  These limits can be obtained by inspection of the expressions for the meson propagators (for details see \cite{RuivoSousa,costaruivo}).
We do not show the corresponding limits for the pions because, in the range of densities studied, the pion modes  remain outside the continuum.

Concerning the Dirac sea excitations, we observe the expected splitting between charge multiplets: the increase of the masses of $K^+ ,K^0$ and $\pi^- $ with respect to those of $K^-, \bar K^0 $ and $\pi^+$, respectively, is due to Pauli blocking. 
At the critical density the antikaons enter in the continuum, but, in matter with $\beta$ --equilibrium, they become again bound states above $\rho_0 \sim 4 \rho_0$. The difference between the behavior of mesons in matter with and without strange valence quarks (the right and left panels of Figs. 6 and 7, respectively) is more evident for kaons than for pions, as expected, since the strange degree of freedom for pions only contributes explicitly through the projector $P_{ab}$ [see (\ref{pab})], and for kaons it contributes through its quark structure. The dominant effect is the reduction of the splitting between the kaon and antikaon masses, which is a combination of many--body effects and restoration of chiral symmetry (Fig. 7).

\begin{figure}[t]
\begin{center}
\epsfig{file=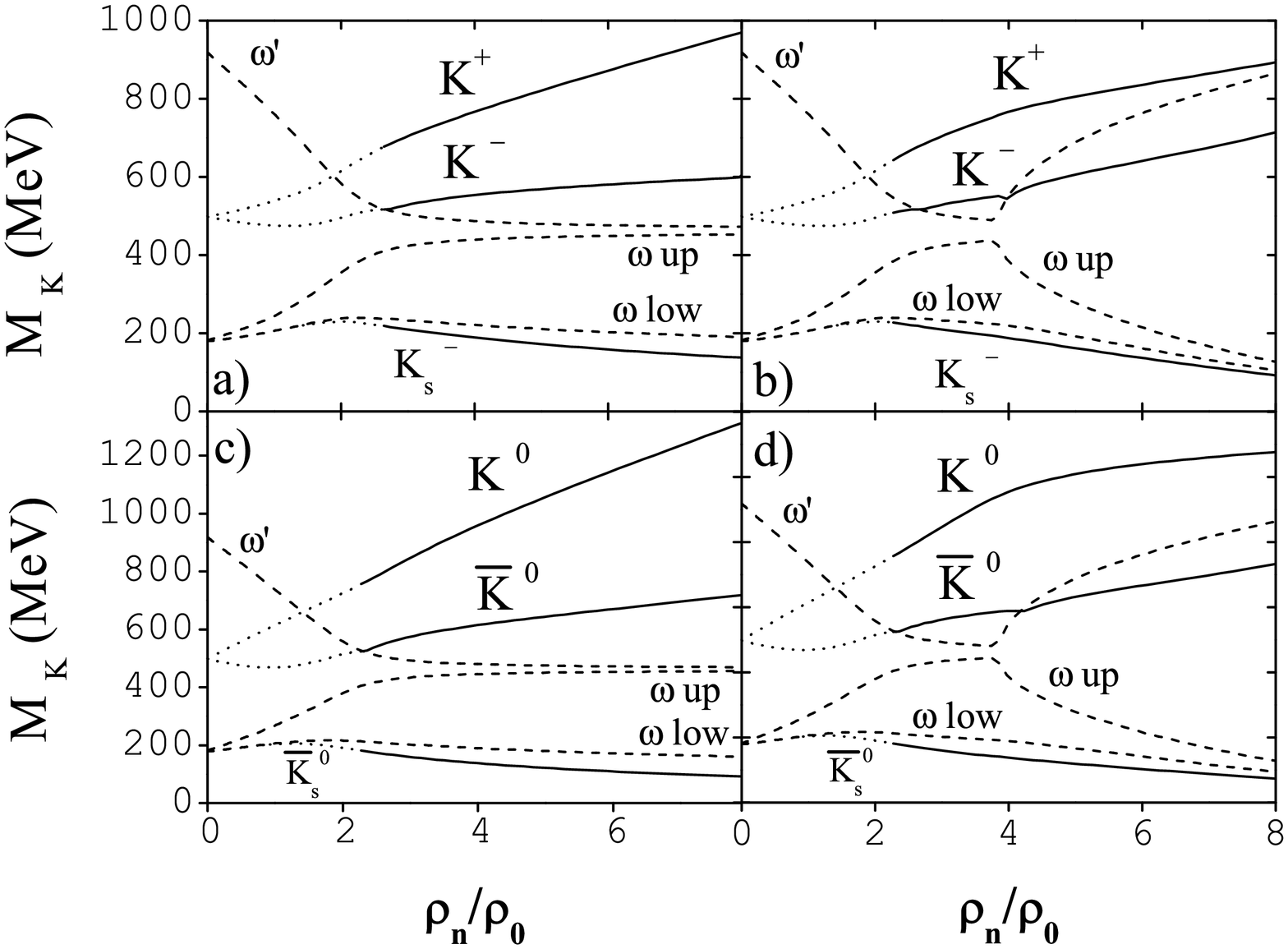, width=16.0cm,height=13cm}
\end{center}
\caption{Kaon and antikaon masses as function of density at $T=0$:
Case I (a) and c)) and  with $\beta$ --equilibrium (b) and d)). $\omega$' -lower limit of the Dirac sea continuum;  $\omega_{up}\,,\omega_{low}$ - limits of the Fermi sea continuum.}
\end{figure}

Let's now analyze the other type of solution, which we denoted by the subscript $S$.
Below the lower limit of the Fermi sea continuum of particle--hole excitations there are low bound states  with quantum numbers of $K^-\,,\bar K^0\,,\pi^+$, respectively. These low-energy modes are associated to a first-order phase transition. As a matter of fact, it was shown in \cite{RSP,costa} that in the NJL model including vector mesons, where a crossover occurs, the splitting between the excitations of the Dirac sea is more pronounced and the low-energy solutions do not appear. 
A general conclusion from our exploration of the high-density region, whether or not we have matter in $\beta$ --equilibrium, is that the low-energy modes in this region are poorly collective and, therefore, have little
strength, contrary to what happens in the low-density region.
The pion low-energy mode merges even in Fermi sea continuum just after $\rho_n=\rho_n^{cr}$ \cite{costaruivo}.  

Since the strangeness degree of freedom is more relevant for kaons than for pions, we will analyze the results for kaons in more detail, in particular  discussing what happens in Cases II and III. Complementary information concerning the results shown in Figs. 6-7 is given by the quark-meson coupling constants displayed in Fig. 8. The coupling constant for kaons decreases with density, which is consistent with the increasing of its mass; for antikaons, it remains constant when the modes are in the continuum, but, if they get outside of the continuum it increases again, which is consistent with the interpretation given above that they are bound states in that region of densities. Concerning the low-energy antikaon, it becomes less bound as the density increases.

In Fig. 9 the results for the masses and meson-quark coupling constants are shown for Cases II and III. In Case III (matter completely flavor symmetric) all the charged and neutral kaons and antikaons are degenerated, as expected; its mass increases and its coupling constant decreases accordingly.
The low-energy antikaon does not exist, which is in agreement with its origin: a particle-hole excitation due to the flavor asymmetry of matter. The behavior for Case II is qualitatively similar to matter in $\beta $ --equilibrium, which is natural, since the strange valence quarks only appear at some density. The main differences are that the neutral kaons are
degenerated with the charged ones, due to isospin symmetry, and the upper antikaon gets out the continuum at a higher density than in the case of $\beta$ --equilibrium and is less bound, which is due to later appearance and smaller fraction of strange valence quarks, in this case.

\begin{figure}[t]
	\begin{center}
		\epsfig{file=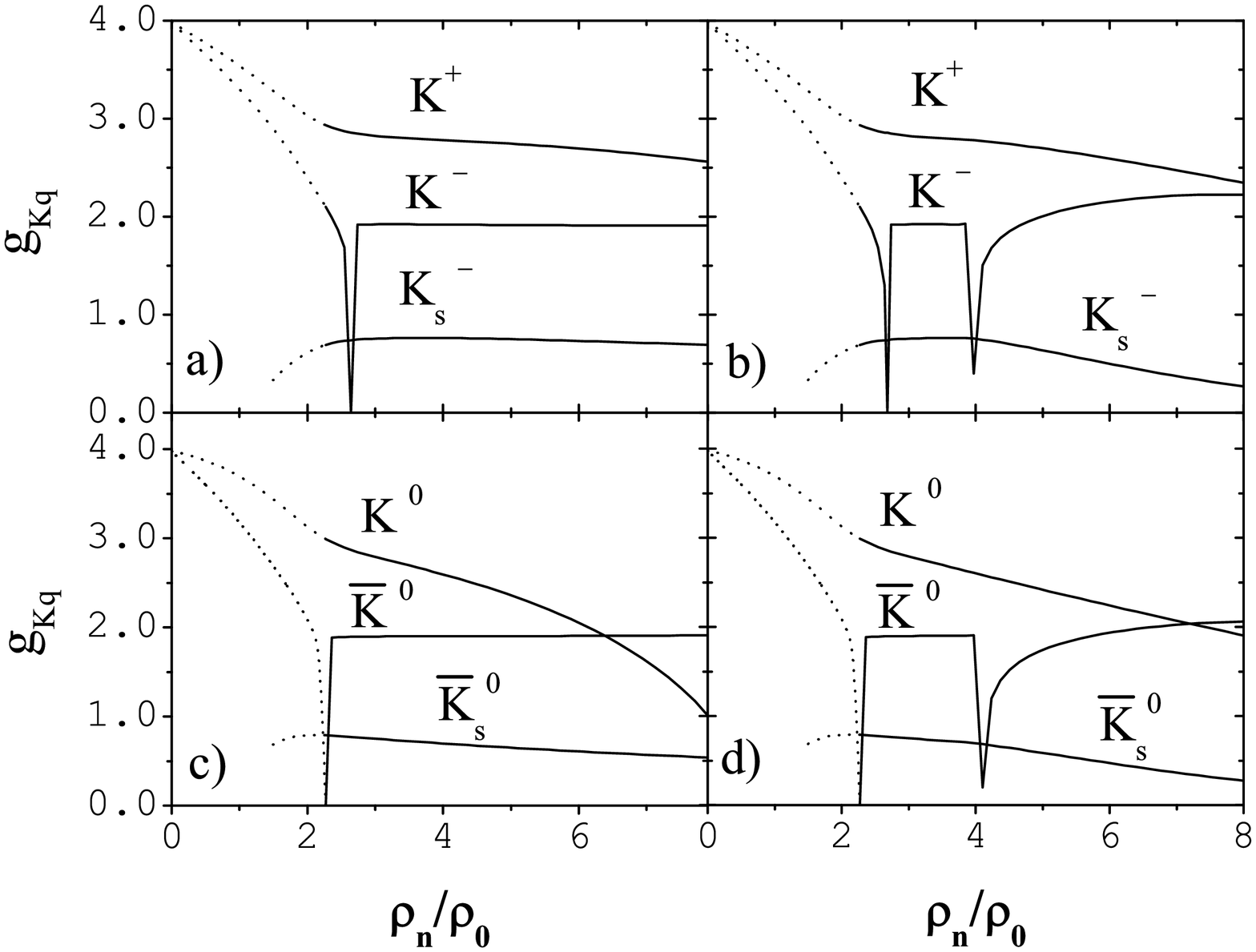, width=16.0cm,height=13cm}
	\end{center}
	\caption{Density dependence of meson-quark coupling constants for $T=0$: 		Case I a) and c); with $\beta$ --equilibrium b) and c).}
\end{figure}
Finally, some remarks are in order concerning our results.  Below the critical density  the system is in a mixed phase, therefore, there is no clear definition of hadron masses, that is why we plotted the masses as dotted lines in this region (they should be understood as average values, meaningful only from a qualitative point of view).  Moreover, it is clear that a Fermi sea of quarks is certainly not  a good description of nuclear matter in the confined phase, and the results in that region within NJL models should be taken with caution. The  results for the high-density region are more reliable, since at such densities the quarks are supposed to be deconfined and a Fermi sea of quarks is a reasonable description of matter in that region. 

A question that can be naturally raised is if our model allows for the possibility of kaon condensation.  The idea of kaon condensation and the recognition of its relevance for astrophysics have been explored since the 80's \cite{Kaplan,kubodera,lee}. In our present approach, however, the criterion for the occurrence of kaon condensation is not satisfied since the antikaon masses   are always larger than the difference between the chemical potential of strange and nonstrange quarks. Our results should be understood as a starting point for further refinements, since it is known that in matter of high-density and low temperature the system might undergo one  or more phase transitions.
For instance, we have not taken into account the effect of color superconductivity that is expected to provide an extra binding mechanism. This mechanism is more important for strange quarks, leading to the CFL phase \cite {CFL,rajagopal}, with kaon condensates, and is supposed to have  interesting consequences for the structure of neutron stars. 
This is out of the scope of the present paper, but work in this direction is in progress.
\begin{figure}[t]
\begin{center}
\hspace*{-0.2cm}\epsfig{file=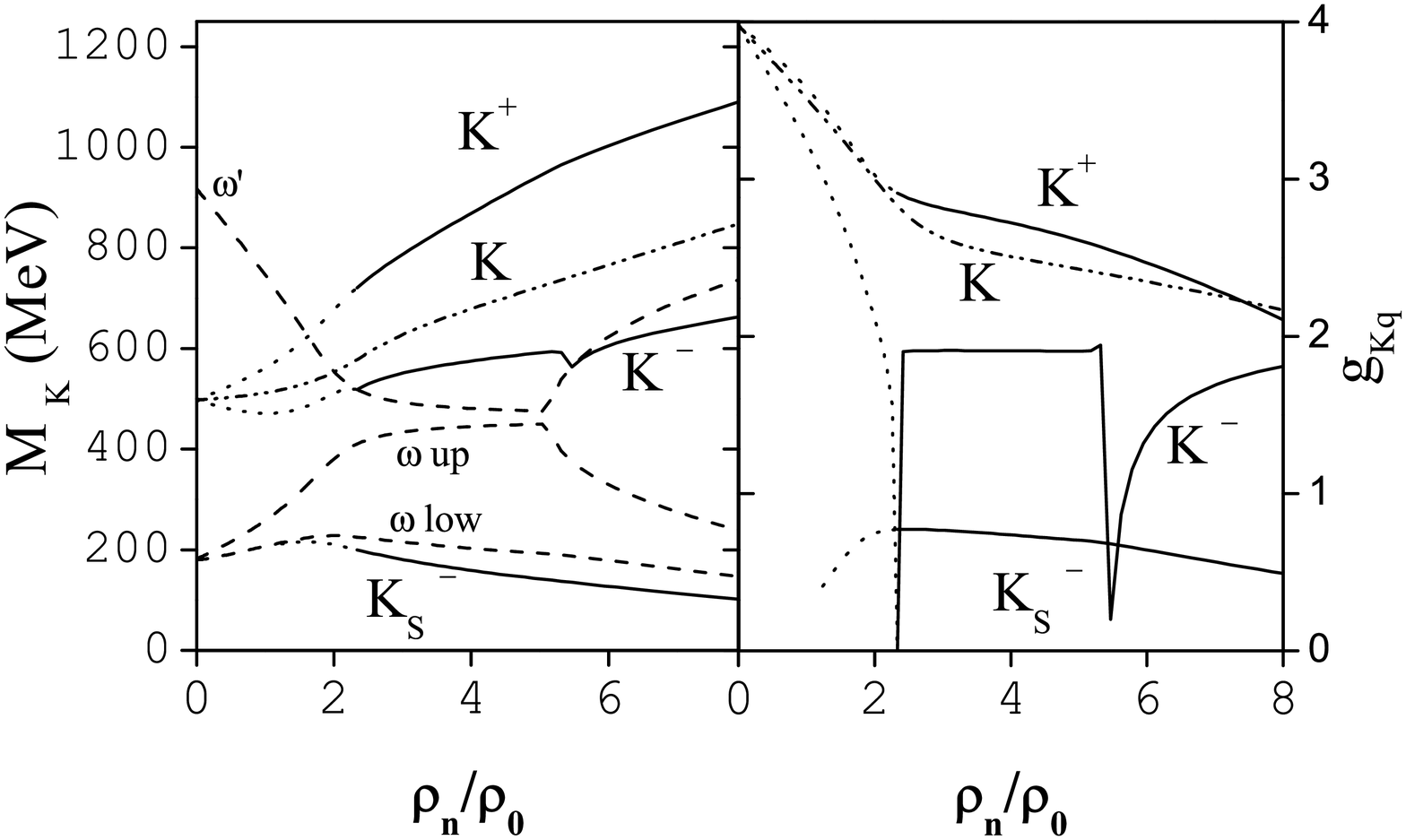, width=15.0cm,height=9cm}
\end{center}
\caption{Kaon and antikaon masses (left panel) and meson-quark coupling 
constants (right panel) as function of density for Case II and Case III at $T=0$. 
In Case II $K^0$, $\bar{K}^0$ and $\bar{K}_S^0$ are degenerated with $K^+$, $K^-$ and $K_S^-$ respectively.
In Case III all kaons are degenerated (dash-dot-dot line).}
\end{figure}


\subsection{Behavior of $\eta\, \mbox{and}\, \eta'$}
 
Now, we analyze the results  for the masses of $\eta$ and $\eta'$.
As we can see, the $\eta'$ --meson lies above the quark--antiquark threshold for $\rho_n < 2.5 \rho_0$ and it is a resonant state. 
After that density, the $\eta'$ becomes a bound state. This is due to the fact that the limits of the Dirac sea increase with density ($\omega_{u,d}=2 \mu_{u,d}$ and $\omega_{s}=\sqrt{M_s^2+\lambda_s^2}$ at $\rho\not= 0$, instead of $\omega_i=2 M_i$ at $\rho=0$).
 
Interest in the study of in-medium properties of $\eta, \eta'$ is in part motivated by the conjecture that their behavior could give indications of possible restoration of the $U_A(1)$ symmetry, in particular their masses could eventually be degenerated. We will show that within our model this does not happen, the behavior of  $\eta\,\mbox{and}\, \eta'$  essentially reflects the tendency to restoration of  chiral symmetry in different sectors.

It has been argued that, in principle, there is no reason why the parameters used to fix the model in vacuum should not depend on density or temperature. In \cite{kuni} a dependence on temperature of the coupling constant ($g_D(T)= g_D(0) $exp$[-(T/T_0)^2]$) was used in order to investigate the restoration with temperature of the $U_A(1)$ symmetry, that is explicitly broken in this model by the 't Hooft interaction. In the present calculation our parameters, including $g_D$ are kept constant, so it would be expected that  no indications of restoration of $U_A(1)$ symmetry would be found.

The masses, plotted in Fig. 10, exhibit a tendency to level crossing but, after the critical density, they split again, the splitting being more pronounced in the case of matter without strange quarks. This is  due to the absence of strange valence quarks in this  case and is related with the in--medium behavior of the mixing angle, $\theta$, between $\eta$ and $\eta'$.   
\begin{figure}[t]
\begin{center}
\hspace*{-0.2cm}\epsfig{file=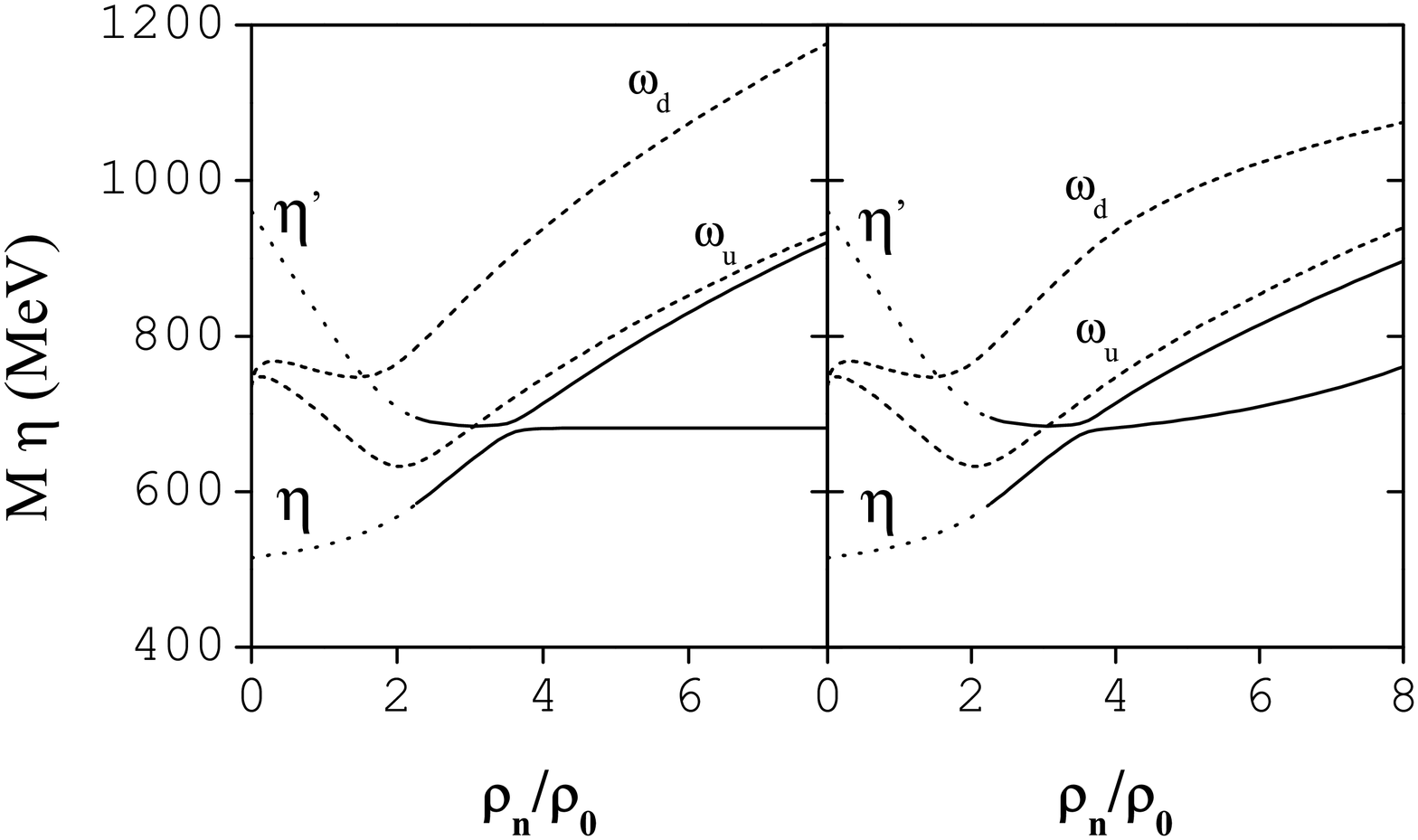, width=14.0cm,height=9cm}
\end{center}
\caption{
$\eta$ and $\eta'$ masses as function of density without (left panel) 
and  with (right panel) $\beta$ --equilibrium at $T=0$.  $\omega_{u}\,,\omega_{d}$ - limits of the Dirac continua.}
\end{figure}
As it was shown in  \cite{neutral}, above $\rho\simeq 3.5 \rho_0$ the angle becomes positive and increases rapidly; a similar behavior with temperature was found in the framework of the $\sigma$ model \cite{sigma}. Therefore, the strange quark content  of the mesons changes: at low-density, the $\eta '$ is more strange than  the $\eta$ but the opposite occurs at high-density. 
So, in a medium without strange quarks the $\eta$ mass should stay constant in the region where its content is dominated by the strange quark. 

Finally, we add some remarks on the approximations made in the study of $\eta\,,\eta'$ in--medium. Let us focus again on the projector $P_{ab}$ and $\Pi_{ab}$ (\ref{Pab}).
The nondiagonal elements that describe the mixing of $\pi^0\,-\,\eta$ and of $\pi^0\,-\,\eta'$ are proportional to $<\bar{q}_{u}\,q_u>\,-<\bar{q}_{d}\,q_d>$ for $P_{ab}$ and to $J_{uu}(P_0)-J_{dd}(P_0)$ for $\Pi_{ab}$. In the cases here considered of matter with isospin asymmetry (matter in $\beta$ --equilibrium and Case I) these quantities are nonzero. Here we did the approximation of neglecting them and we checked, by means of a simple estimation, that this is a reasonable approximation.


\section{Mesons in hot and dense matter}\label{hotand}
 
As it has been discussed in Section IV, the phase transition in hot and dense matter is of first-order below $T_{cl}\sim 56$ MeV. Above $T_{cl}$ we have a crossover. We define the critical point as the temperature and density where the pion becomes unbound (Mott transition point). Since the  model has no confinement, the system is unstable against expansion, but, in  spite of these drawbacks, we think it is illustrative to plot the meson masses as functions of temperature and density. We consider here only the case of neutron matter in $\beta$ --equilibrium.  
As already discussed in Section \ref{secpion}, in order to investigate  possible restoration of symmetries,  it is important to   study the mesons that have  phenomena of  symmetry  breaking at their origin, whether it is chiral or $U_A(1)$ symmetry.  As it will be shown, the dominant effect found in our results is   the restoration of chiral symmetry.  

A first conclusion is that there are differences in  the mesonic behavior as compared  to the zero temperature case discussed in last section, where we have seen that some mesons are still bound states in the chiral restored phase. 
Here, similarly to finite temperature and zero density case, the $\bar q q$ threshold for the different mesons is at the sum of the constituent quark masses, so the mesons dissociate at densities and temperatures close to the critical ones. 
A second feature to be noticed is that the  mesons that are remnants of  Goldstone bosons show more clearly the difference between the chiral symmetric and asymmetric phases.  
The slight differences of behavior inside each flavor multiplet are due to many--body effects. 
It can be seen in the diagrams that there is a "line" separating the regions of the surface with different curvatures. 
This is very clear for the case of pions (Fig. 11) and kaons (Fig. 12), and, as shown in \cite{neutral}, this is also apparent for the $\eta$. 
We may call this line the "Mott circle", since it separates the region where the meson are bound states from the region where they are in the continuum. 

\begin{figure}[t]
\begin{center}
  \begin{tabular}{cc}
    \epsfig{file=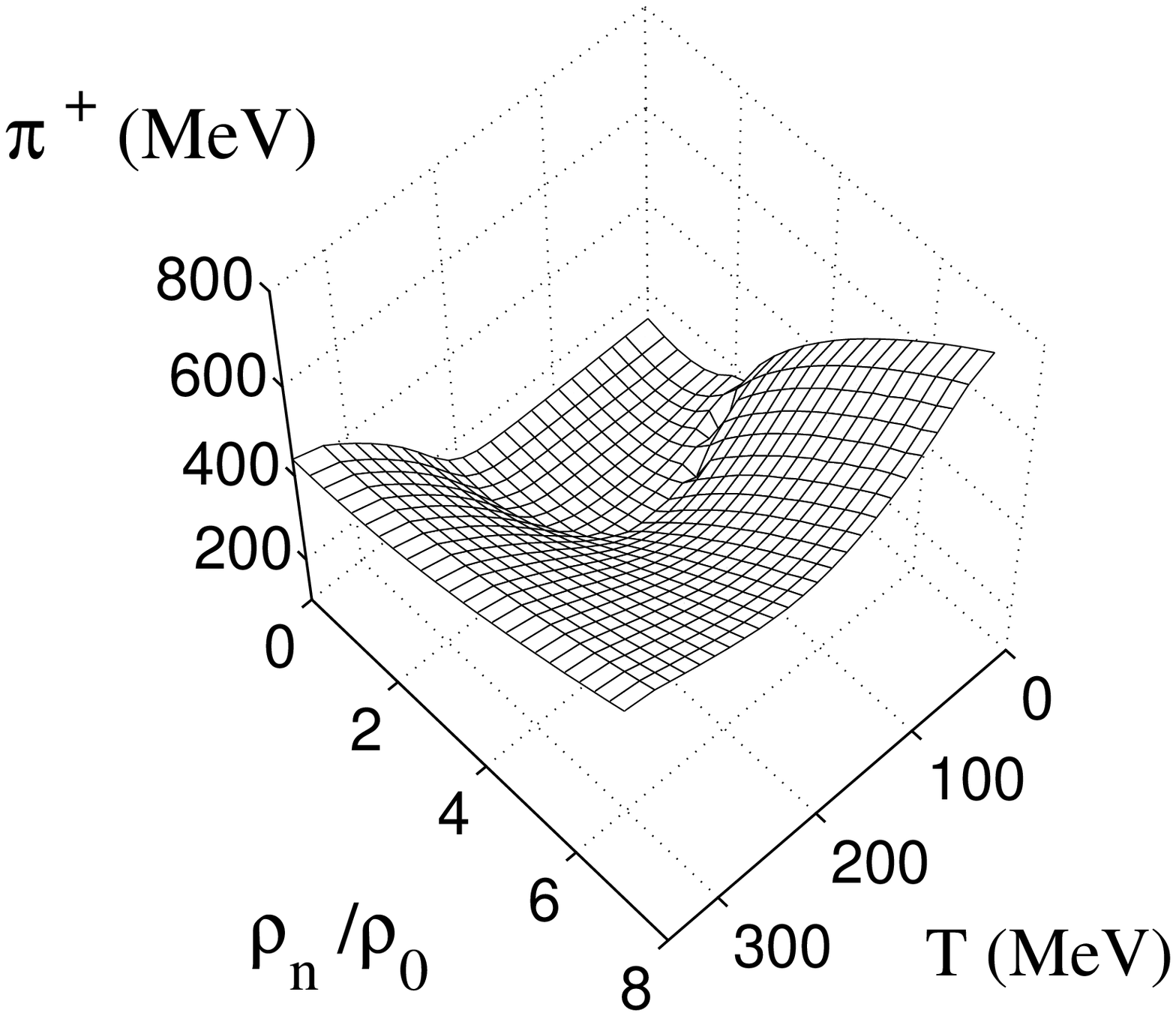,width=8cm,height=6.5cm} &
    \epsfig{file=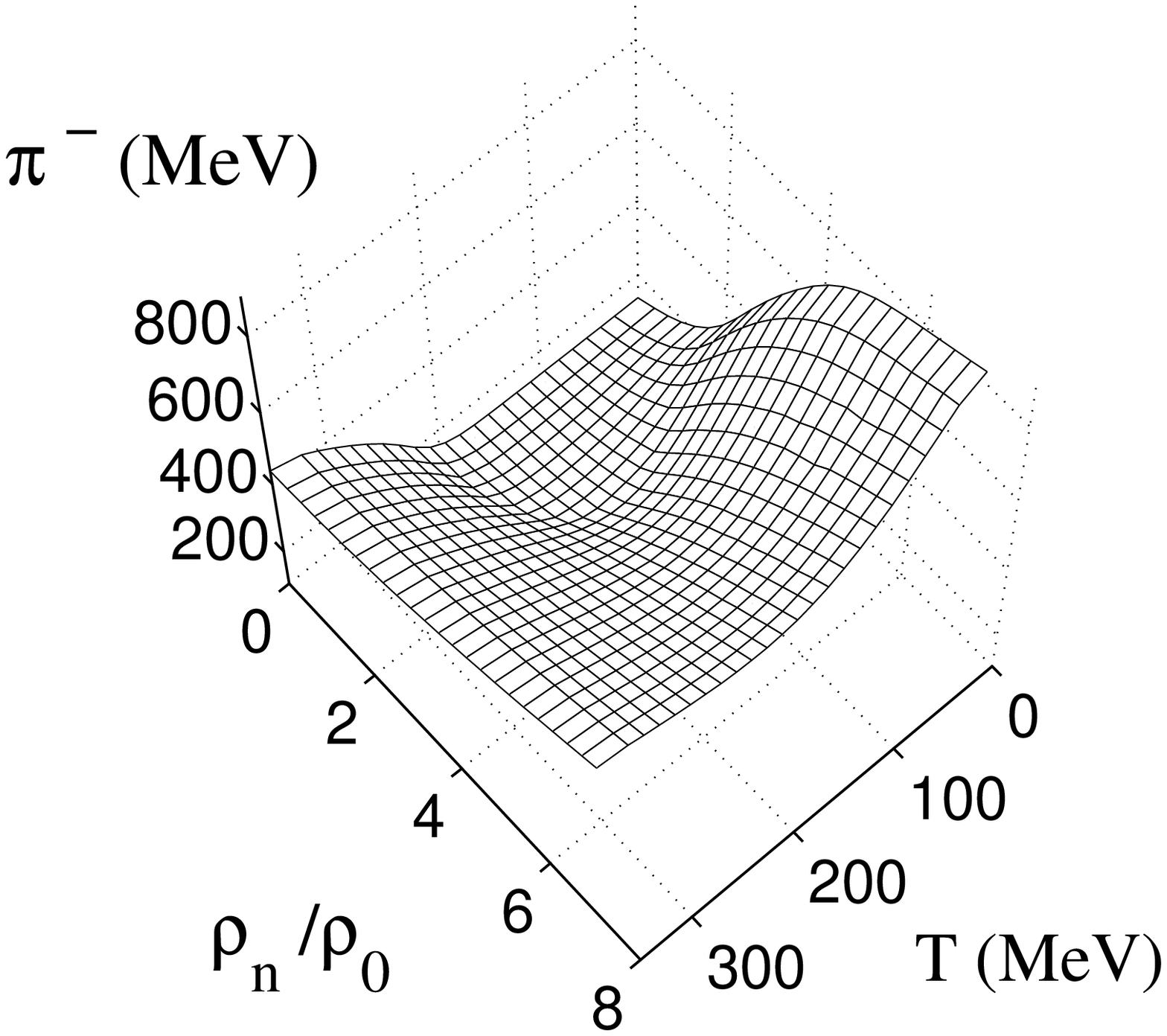,width=8cm,height=6.5cm} \\
  \end{tabular}
  \begin{center}
   	\epsfig{file=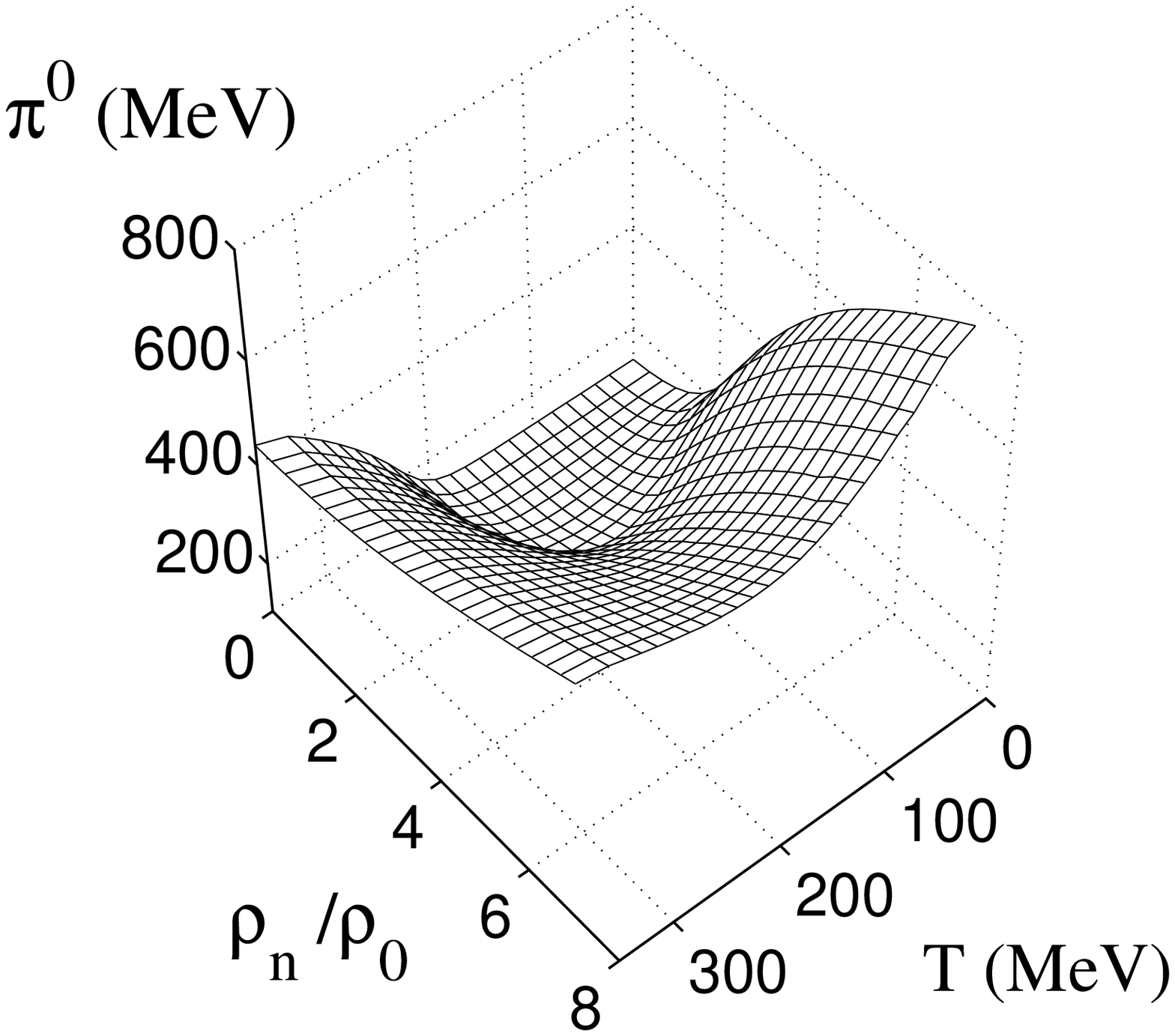,width=8cm,height=6.5cm} 
  \end{center}
\end{center} 
\caption{Pion masses as functions of temperature and density.}
\end{figure}
\begin{figure}[t]
\begin{center}
  \begin{tabular}{cc}
   	\epsfig{file=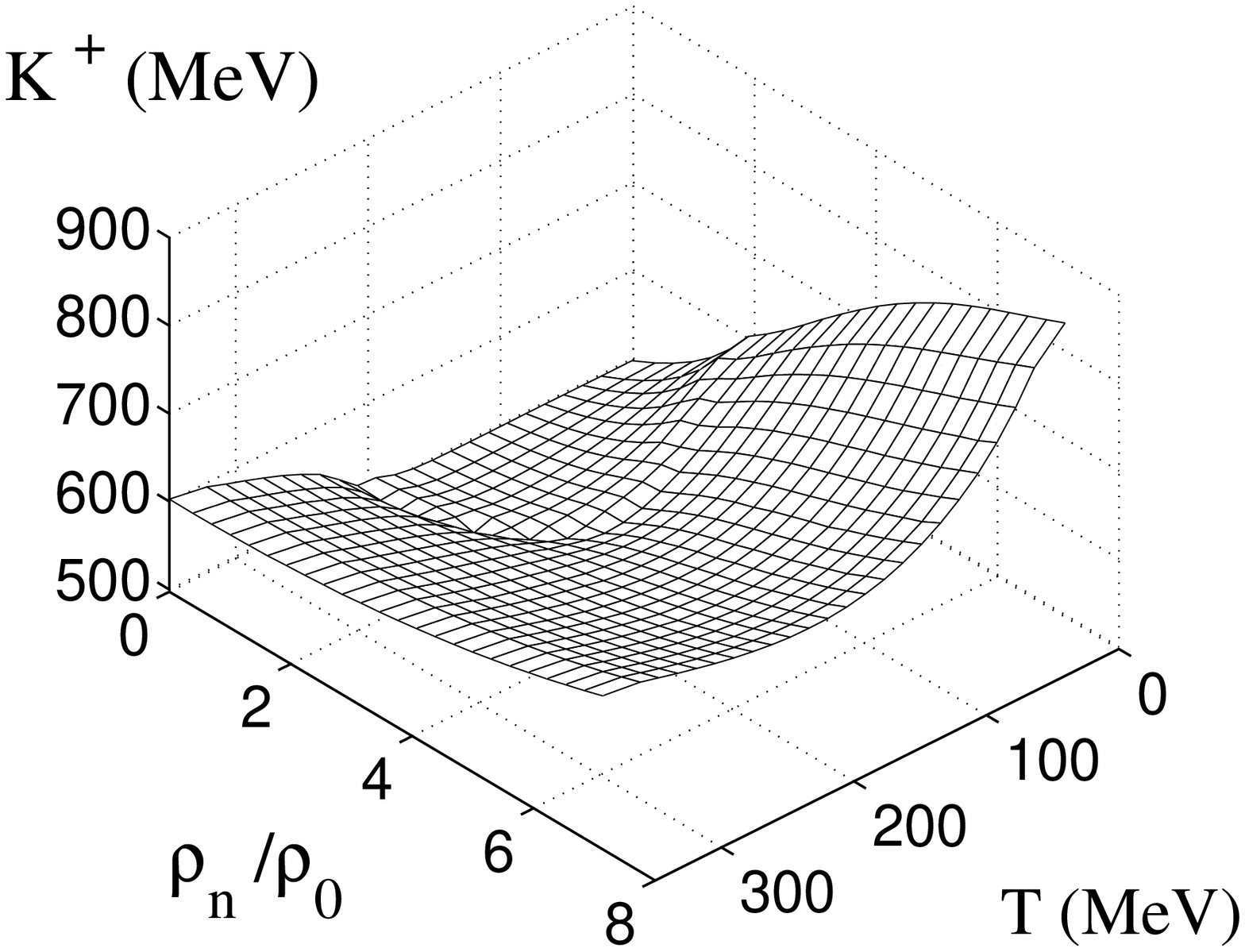,width=8cm,height=7cm} &
   	\epsfig{file=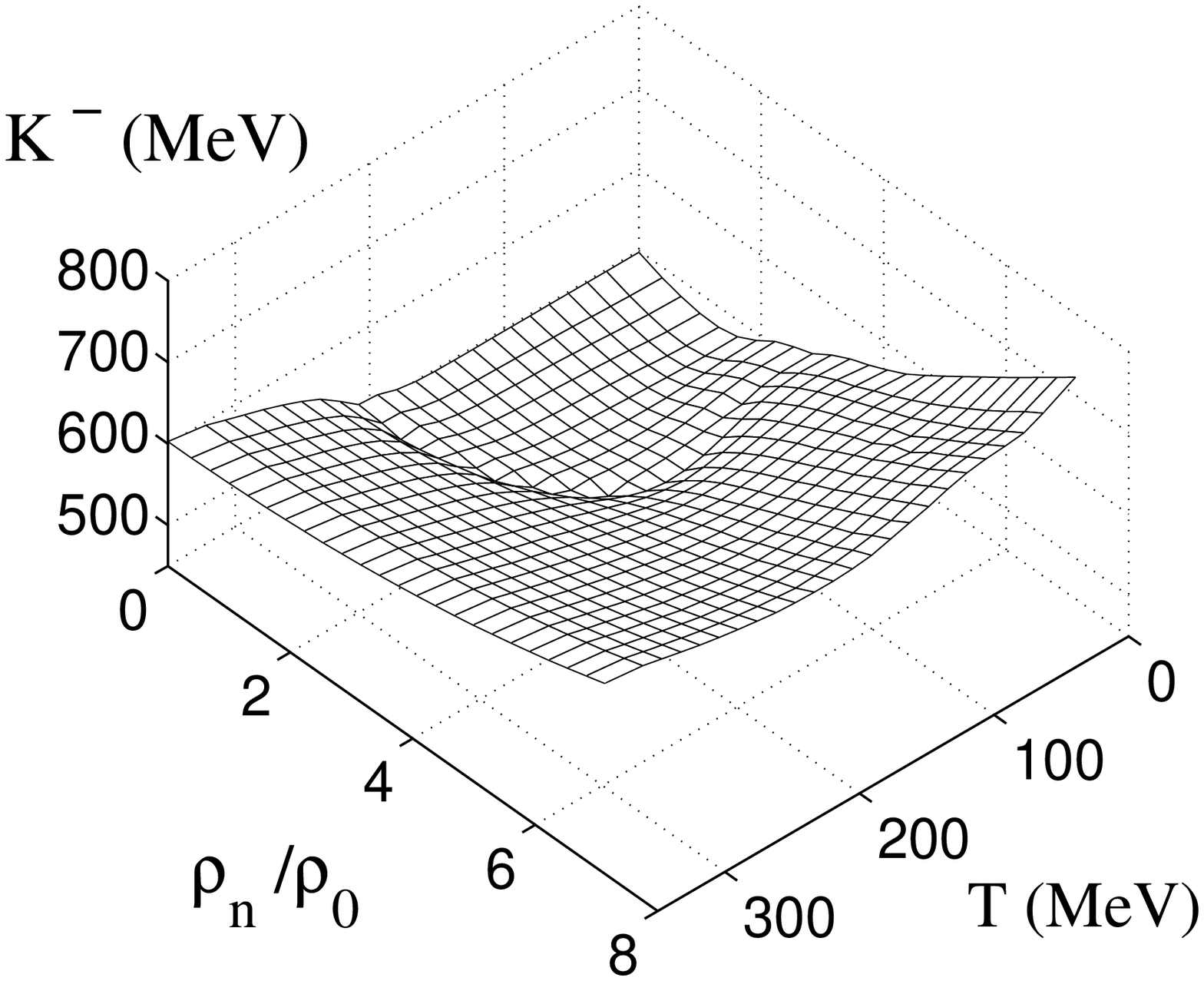,width=8cm,height=7cm} \\
  \end{tabular}
  \begin{tabular}{cc}
   	\epsfig{file=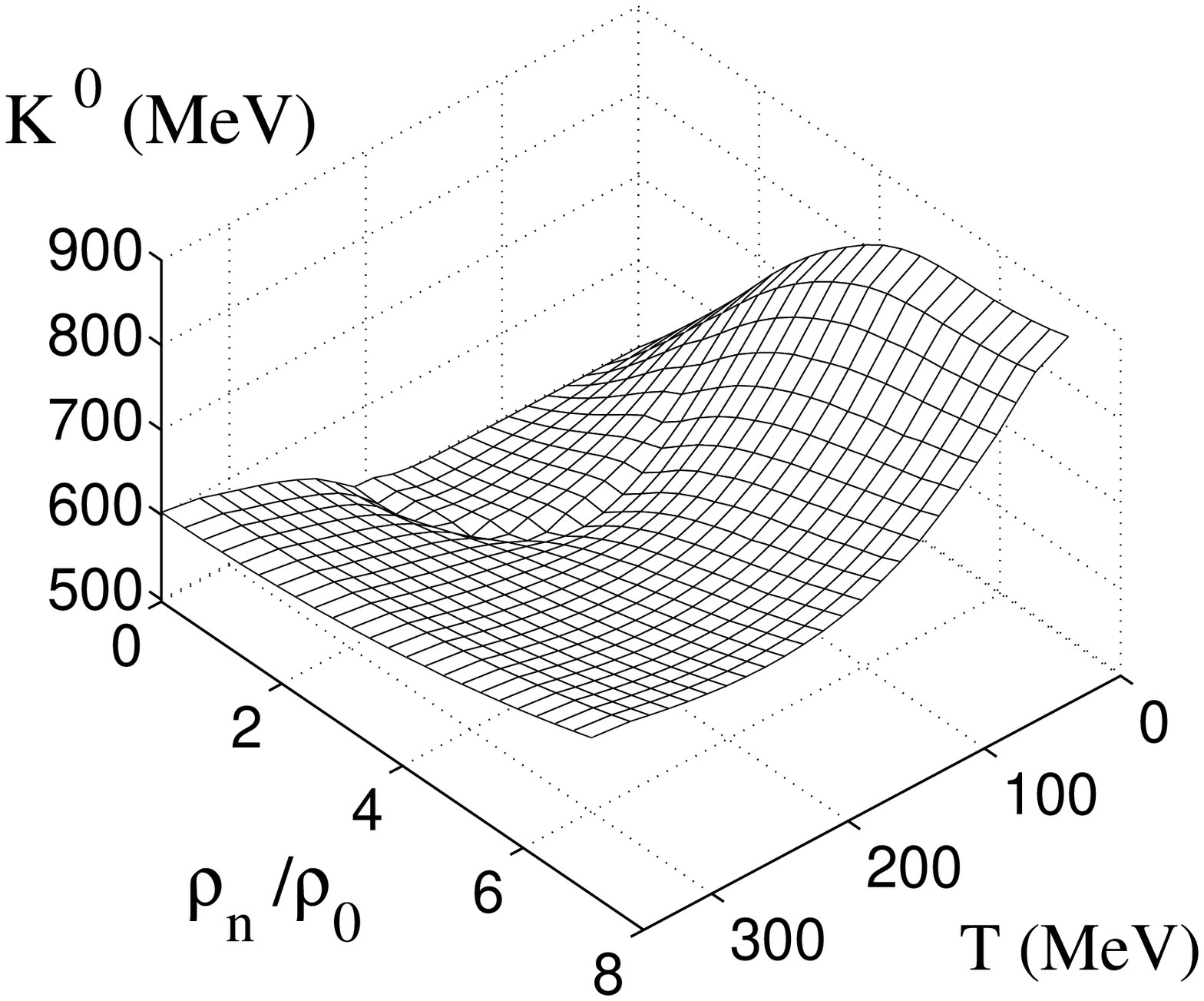,width=8cm,height=7cm} &
    \epsfig{file=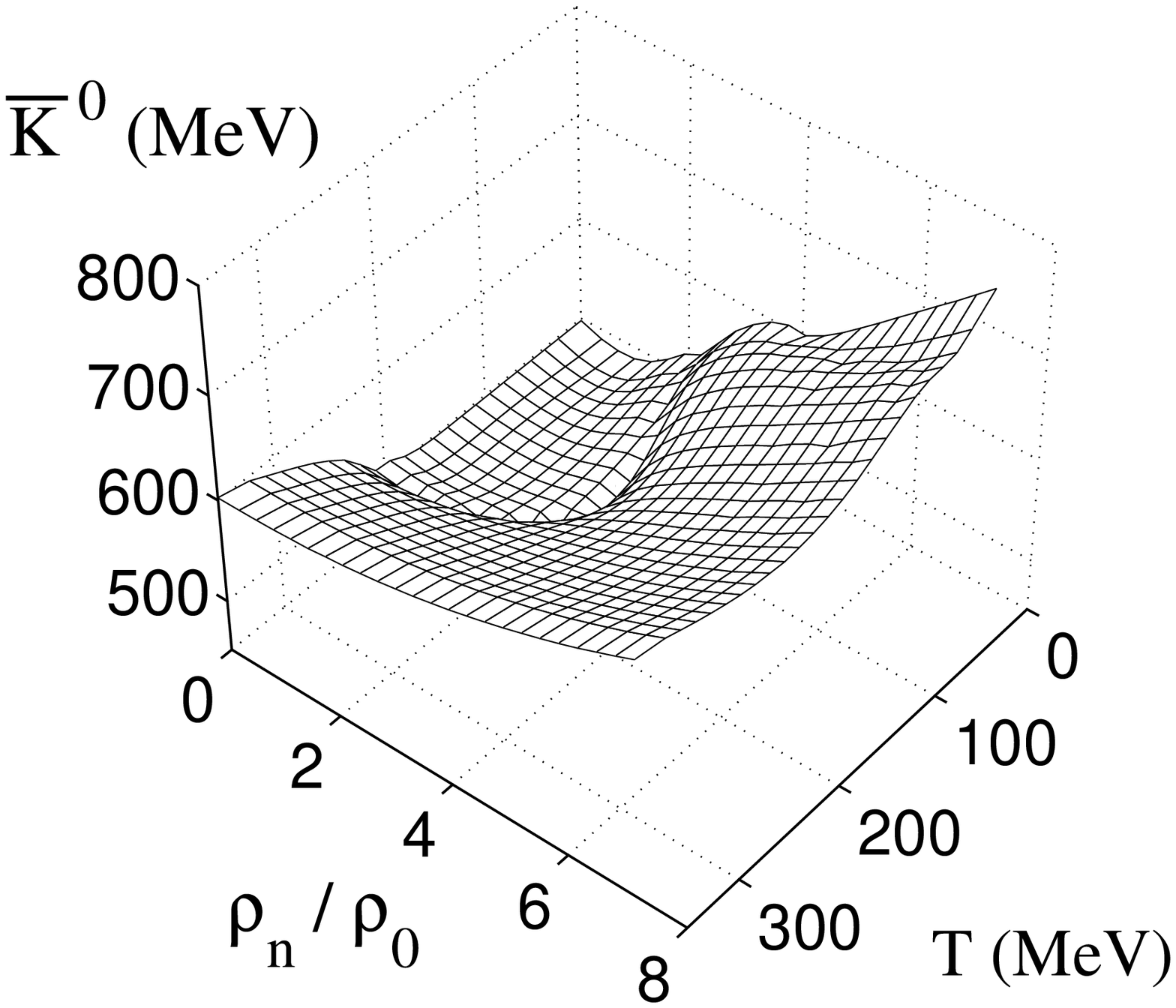,width=8cm,height=7cm} \\
  \end{tabular}
\end{center}
\caption{Kaon masses as functions of temperature and density.} 
\end{figure}

We notice that in the context of our model, Mott transition is certainly related with the chiral transition but can not be seen as a mechanism for quark deconfinement. 
In fact, the NJL model is suitable to describe the chiral phase transition but not the deconfinement phase transition, since it has no confining mechanism. However, in spite of this drawback, the model gives a reasonable description of a quark phase, at high-densities and temperatures, where interacting quarks with small constituent mass are supposed to exist. In this phase the mesons are unstable resonances, decaying in pairs of quark--antiquark states.

The behavior of the $\eta'$ is more involved since this meson is not a Goldstone boson associated with spontaneous chiral symmetry breaking, 
and, in the present model is described as a $\bar q q $ resonance, even at zero temperature and density (see Fig. 10). Only at zero temperature and for $\rho_n > 3.0 \rho_0$ it becomes a bound state (for more details see \cite{neutral}).

Finally, we notice that, as soon as the system heats, and since very low temperatures, the low-energy modes stay  inside  the low continuum and, unlike the zero temperature   case, they do not become bound states.


\section{Summary and conclusions}\label{conclusion}
 
To conclude and summarize, in the present paper we have investigated phase transitions in hot and dense matter and the  in--medium behavior of the 
pseudoscalar mesons, in the framework of the $SU(3)$ Nambu--Jona-Lasinio
model, including the 't Hooft interaction, which breaks the $U_A(1)$ symmetry.

Three scenarios were considered: i) zero chemical potential and finite temperature, ii) zero temperature and finite chemical potential in four types of quark matter and, finally, iii) finite temperature and density.
Although in all the cases we found (partial) restoration of chiral symmetry, different features occur in several observables.

Concerning case i) we mainly reproduce results obtained by other authors \cite{njlT}, in order to allow for comparison with situations ii) and iii) that represent our original contribution. The main feature is the dissociation of mesons at the Mott transition point that occurs when the meson masses equal the sum of the masses of their constituents. After that point, which sets the corresponding threshold of the $\bar q q $ continuum, the mesons cease to be bound states and become resonances. 

New features occur at zero temperature and finite density, in flavor asymmetric medium (ii)).
In order to discuss the nature of the phase transition, we plot the pressure and energy per particle versus baryonic density. 
For a suitable choice of the parameters we have a mixed phase, which may be
interpreted as the system having a hadronic phase with partially restored chiral symmetry embedded in a nontrivial vacuum. After the critical density we have a quark phase. 
We notice that in flavor asymmetric matter the energy of the stable hadronic phase, at $\rho_{cr}$, is in a region where strange valence quarks are still absent, so SQM only could exist in a metastable state. Only for the case of equal number of quarks $u\,,d\,,s$ we found stable SQM, but with a higher energy per particle then atomic nuclei. Concerning the masses of the mesons, there is a splitting between the flavor multiplets in flavor asymmetric matter and, for kaons and pions, low-energy modes with quantum numbers of $\pi^+, K^-$ and ${\bar K}^0$ appear. 
Our results for kaons in flavor asymmetric matter show that the splitting between kaons and antikaons increases with density, which is compatible with experimental results \cite{Herrman} that indicate a reduction (enhancement) of kaon (antikaon) production in medium. In the high-density region the splitting has a different behavior according to whether there are strange quarks present or not. However, although the splitting is reduced in matter with strangeness (the smaller splitting is in matter in $\beta$ --equilibrium where there is a larger fraction of strange quarks), the upper antikaon becomes more bound in this case, so it is likely that the splitting can also be observed.
The lower antikaons are more bound in the low-density region; these modes do not exist in flavor symmetric matter. 
Although we find kaons as  bound states in the high-density region, we do not find kaon condensation. 
Concerning $\eta $ and $\eta'$, their masses come closer up to the critical density, but after that point they split again, the splitting being more pronounced in the case of neutron matter without $\beta$  --equilibrium, which is related with the change of the strangeness content of the mesons. 

In situation iii), hot and dense matter, the phase transition becomes a crossover above the critical "end point" $T= 56$ MeV, $\rho=1.53 \rho_0$, the system having a mixed phase before that point.  In this case, the $\bar q q $ continuum turns again to be at the sum of the constituent quark masses.
By plotting the meson masses we see that there is a line along which the
curvature of the surfaces changes, indicating the partial restoration of chiral symmetry. 
Beyond that line,  the mesons become resonances.


\begin{acknowledgments}
We are grateful to the referee for valuable comments and criticisms which contributed for the improvement of this paper.   
Work supported by grant SFRH/BD/3296/2000 (P. Costa), Centro de 
F\'{\i}sica Te\'orica, FCT, GTAE and RFBR 03-01-00657 (Yu. L. Kalinovsky).

\end{acknowledgments}


\appendix \label{apendice}
\section{}

In this appendix we give some details about the calculation of the effective action and of the integrals appearing in the meson propagators, in the vacuum and at finite temperature and density.


\subsection{Calculation of the effective action}

 The   model Lagrangian (\ref{lagr})  can be put in a form suitable for the usual bosonization procedure after an adequate treatment of the last term that contains a  six quark interaction. 
To obtain a four quark interaction  from the six quark interaction we make a shift
$(\bar{q} \lambda^a q) \longrightarrow (\bar{q} \lambda^a q) + <\bar{q} \lambda^a q>$,
where $<\bar{q} \lambda^a q>$ is the vacuum expectation value, and
contract one bilinear $(\bar{q} \lambda^a q)$. Then the following effective quark Lagrangian is obtained:
\begin{eqnarray}
{\cal L}_{eff} &=& \bar q\,(\,i\, {\gamma}^{\mu}\,\partial_\mu\,-\,\hat m)\, q \,\,
\nonumber \\
&+& S_{ab}[\,(\,\bar q\,\lambda^a\, q\,)(\bar q\,\lambda^b\, q\,)]
+\,P_{ab}[(\,\bar q \,i\,\gamma_5\,\lambda^a\, q\,)\,(\,\bar q
\,i\,\gamma_5\,\lambda^b\, q\,)\,],
\label{lagr_eff}
\end{eqnarray}
were  the projectors $S_{ab}\,, P_{ab}$ are of the form:
\begin{eqnarray}
S_{ab} &=& g_S \delta_{ab} + g_D D_{abc}<\bar{q} \lambda^c q>, \label{sab}\\
P_{ab} &=& g_S \delta_{ab} - g_D D_{abc}<\bar{q} \lambda^c q>. \label{pab}
\end{eqnarray}
The constants $D_{abc}$ coincide with the $SU(3)$ structure constants $d_{abc}\,\,$ for
$a,b,c =(1,2,\ldots ,8)$ and $D_{0ab}=-\frac{1}{\sqrt{6}}\delta_{ab}$, $D_{000}=\sqrt{\frac{2}{3}}$.
The hadronization procedure can be done by the integration over the quark fields
in the functional integral with (\ref{lagr_eff}), leading to the effective action (\ref{act}).


\subsection{Integrals}

The polarization operator in Eq. (5) in the momentum space has the form,
\begin{eqnarray}\label{polop}
\Pi_{ab} (P) = i N_c \int \frac{d^4p}{(2\pi)^4}\mbox{tr}_{D}\left[
S_i (p) (\lambda^a)_{ij} (i \gamma_5 )
S_j (p+P)(\lambda^b)_{ji} (i \gamma_5 )
\right],
\end{eqnarray}
where $\mbox{tr}_{D}$ is the trace over Dirac matrices.

In the expressions of $\Pi^{ij}$ in Eq. (5), at $T=0$ and $\rho=0$, we have used the following integrals:

\begin{eqnarray}\label{i1}
I_1^i &=& i N_c \int \frac{d^4p}{(2\pi)^4} \, \frac{1}{p^2-M_i^2}
       = \frac{N_c}{4 \pi^2} \int^{\Lambda}_0 \frac{{\tt p}^2 d {\tt p}}{E_i} ,
\end{eqnarray}
\begin{eqnarray}\label{i2}
I_2^{ij}(P_0) &=& i N_c \int \frac{d^4p}{(2\pi)^4} \, \frac{1}{(p^2-M_i^2)((p+P_0)^2-M_j^2)}
\nonumber \\
       &=& \frac{N_c}{4 \pi^2} \int^{\Lambda}_0 \frac{{\tt p}^2 d {\tt p}}{E_i E_j}
       \,\,\, \frac{E_i+E_j}{P_0^2-(E_i+ E_j)^2} \, ,
\end{eqnarray}
where $E_{i,j}=\sqrt{{\tt p}^2+M_{i,j}^2}$ is the quark energy.
To regularize the integrals we introduce the 3--dimensional cutoff
parameter $\Lambda$. When $P_0 > M_i+M_j$ it is necessary  to take into account
the imaginary part of the second integral.
It may be found, with help of the $i \epsilon$ --prescription
$P_0^2 \rightarrow P_0^2 - i \epsilon$, that
\begin{eqnarray}\label{ima}
I_2^{ij}(P_0)
       = \frac{N_c}{4 \pi^2}
       {\mathcal{P}}\int^{\Lambda}_0 \frac{{\tt p}^2 d {\tt p}}{E_i E_j}
       \,\,\frac{E_i+E_j}{P_0^2-(E_i+ E_j)^2}
+       i \frac{N_c}{16\pi} \, \frac{p^*}{(E_i^*+E_j^*)}
\end{eqnarray}
with the momentum: $p^*=\sqrt{(P_0^2-(M_i-M_j)^2)(P_0^2-(M_i+M_j)^2)}/2P_0$ and
the energy: $E^*_{i,j}=\sqrt{(p^*)^2+M_{i,j}^2}$.

Concerning the calculation of the propagators of the diagonal  mesons $\pi^0$, $\eta$ and $\eta'$,  the   projector $P_{ab}$ and the polarization operator   $\Pi_{ab}$ (see Eq. (\ref{Pab})) are matrices that, in the case $<\bar{q}_{u}\,q_{u}>=<\bar{q}_{d}\,q_{d}>$, have the  nonvanishing elements:

\begin{eqnarray}
P_{33}&=& g_S +             g_D <\bar{q}_{s}\,q_s>   , \\
P_{00}&=&g_{S}-\frac{2}{3}g_{D}\left( <\bar{q}_{u}\,q_{u}>+<\bar{q}_{d}\,q_{d}>+<\bar{q}_{s}\,q_{s}>\right) , \\
P_{88}&=&g_{S}+\frac{1}{3}g_{D}\left( 2<\bar{q}_{u}\,q_{u}>+2<\bar{q}_{d}\,q_{d}>-<\bar{q}_{s}\,q_{s}>\right) ,\\
P_{08}&=&P_{80}=\frac{1}{3\sqrt{2}}g_{D}\left( <\bar{q}_{u}\,q_{u}>+<\bar{q}_{d}\,q_{d}>-2<\bar{q}_{s}\,q_{s}>\right),
\end{eqnarray}
and
\begin{eqnarray}
\Pi_{00}(P_0)&=&\frac{2}{3}\left[ J_{uu}(P_0)+J_{dd}(P_0)+J_{ss}(P_0)\right] , \\
\Pi_{88}(P_0)&=&\frac{1}{3}\left[J_{uu}(P_0)+J_{dd}(P_0)+4J_{ss}(P_0)\right],\\
\Pi_{08}(P_0)&=&\Pi_{80}(P_0)=\frac{\sqrt{2}}{3}\left[J_{uu}(P_0)+J_{dd}(P_0)
-2J_{ss}(P_0)\right],
\end{eqnarray}
where 
\begin{equation}
J_{ii}(P_0)=4\left(I_1^i-\frac{P_0^2}{2}I_2^{ii}(P_0)\right).
\end{equation}
At  finite temperature and with two chemical potentials, $\mu_i$ and $\mu_j$, the integrals $I_1^i\,, I_2^{ij}\,, I_2^{ii}$ take the form:

\begin{eqnarray}\label{firstt}
I_1^i = - \frac{N_c}{4\pi^2} \int \frac{{\tt p}^2 d{\tt p}}{E_i} \left(
n^+_i - n^-_i
\right),
\end{eqnarray}

\begin{eqnarray}
I_2^{ij} (P_0,T,\mu_i,\mu_j) &=&
- N_c \int \frac{d^3{\bf p}}{(2\pi)^3}
\Biggl[
\frac{1}{2E_i}
\frac{1}{(E_i+P_0- (\mu_i-\mu_j))^2-E_j^2} \,\, n^+_i \nonumber \\
&& \hspace*{1.3cm}
 - \frac{1}{2E_i}  \frac{1}{(E_i-P_0+ (\mu_i-\mu_j))^2-E_j^2} \,\, n^-_i
\nonumber \\ &&
\hspace*{1.3cm}+ \frac{1}{2E_j}\frac{1}{(E_j-P_0+ (\mu_i-\mu_j))^2-E_i^2}
\,\,  n^+_j  
 \nonumber \\
  && \hspace*{1.3cm}- \frac{1}{2E_j}\frac{1}{(E_j+P_0- (\mu_i-\mu_j))^2-E_i^2}
\,\,  n^-_j 
\Biggr], 
\end{eqnarray}
where  $n_i^{\pm}$ are the Fermi distribution functions (\ref{fermi}), defined in Section III.

For the case $i=j$, with imaginary part, we have the expression
\begin{eqnarray}\label{sint}
I_2^{ii}(P_0, T, \mu_i) =
&& - \frac{N_c}{2\pi^2} {   P} \int \frac{{\tt p}^2 d {\tt p}}{E_i} \,\,
\frac{1}{P_0^2-4 E_i^2} \left( n^+_i - n^-_i\right)
\nonumber \\
&& - i  \frac{N_c}{4\pi} \sqrt{ 1- \frac{4 M_i^2}{P_0^2}  }
\left(n^+_i(\frac{P_0}{2}) - n^-_i (\frac{P_0}{2})\right) \,.
\end{eqnarray}


\end{document}